\documentclass[aps,pra,twocolumn,showpacs,floatfix]{revtex4}

\pdfoutput=1

\usepackage{times}
\usepackage{amsfonts}
\usepackage{amssymb}
\usepackage{amsmath}
\usepackage{graphicx}
\usepackage{bm}
\usepackage{verbatim}

\newcommand \bea {\begin{eqnarray} }
\newcommand \eea {\end{eqnarray}}

\newcommand{\beg}{\begin{equation}}
\newcommand{\en}{\end{equation}}

\newcommand \bel  {\begin{align}}
\newcommand \enl  {\end{align}}

\newcommand{\up}{\uparrow}
\newcommand{\dn}{\downarrow}
\newcommand{\dg}{^\dagger}
\newcommand{\ket}[1]{|#1\rangle}
\newcommand{\bra}[1]{\langle#1|}

\begin{document}

\title{Spectroscopy of the soliton lattice formation in quasi-one-dimensional fermionic superfluids with population imbalance}

\author{Roman M. Lutchyn$^{1,2}$, Maxim Dzero$^{1,3}$ and Victor M. Yakovenko$^{1}$}

\affiliation{$^1$Joint Quantum Institute and Condensed Matter Theory Center, University of Maryland,
College Park, MD 20742-4111, U.S.A.\\
$^2$Microsoft Research, Station Q, Elings Hall, University of California, Santa Barbara, CA 93106,  U.S.A. \\
$^3$Department of Physics, Kent State University, Kent, OH 44242, U.S.A.}

\date{
complied \today}

\begin{abstract}
Motivated by recent experiments in low-dimensional trapped fermionic superfluids, we study a quasi-one-dimensional (quasi-1D) superfluid with a population imbalance between two hyperfine states using an exact mean-field solution for the order parameter.  When an effective ``magnetic field'' exceeds a critical value, the superfluid order parameter develops spatial inhomogeneity in the form of a soliton lattice.  The soliton lattice generates a band of quasiparticle states inside the energy gap, which originate from the Andreev bound states localized at the solitons.  Emergence of the soliton lattice is accompanied by formation of a spin-density wave, with the majority fermions residing at the points in space where the Fulde-Ferrell-Larkin-Ovchinnikov (FFLO) order parameter vanishes.  We discuss possibilities for experimental detection of the quasi-1D FFLO state using elastic and inelastic optical Bragg scattering and radio-frequency spectroscopy.  We show that these measurements can provide necessary information for unambiguous identification of the
spatially-inhomogeneous quasi-1D FFLO state and the soliton lattice formation.
\end{abstract}
\pacs{03.75.Kk, 03.75.Hh, 74.25.Gz, 74.20.Mn}
\maketitle

\section{Introduction}\label{sec:intro}

Superconducting pairing in a system with imbalanced population of two fermion species (typically associated with the up and down spin projections) has been studied for a long time.  Fulde and Ferrell~\cite{FF}, as well as Larkin and Ovchinnikov~\cite{LO}, proposed theoretically that the superconducting order parameter in such a system is spatially inhomogeneous.  Although several candidates for this exotic superconducting state (dubbed in the literature as the FFLO state) have been identified since then~\cite{ErRh4B4, CeCoIn5, Cho'09, Lebed'10, Shimahara, Lortz'07}, still there is no conclusive experimental evidence for the existence of this state in crystalline materials (see review \cite{Zwicknagl-2010}).

Recently, great progress has been made in studying BCS-like pairing between fermionic neutral cold atoms.  In these systems, it is possible to engineer a suitable attractive interaction and control population imbalance of the spin up and down states.  Thus, cold atoms have attracted  a lot of attention for experimental~\cite{imbalance, Rice, Hulet2009} and theoretical~\cite{Casalbuoni2004, Sheehy2006, KunYang2001, Burovski2009,Fuchs2004, Vincent_Liu, Parish2007, Orso2007, Luescher2008, Bolech2009, Feiguin2007, Feiguin2009, Heidrich-Meisner2010, Heidrich-Meisner'10, Tezuka2008, Rizzi2008, Wang2009, Batrouni2008, Casula2008, Torma2008, Torma2010, Loh'10, Trivedi'11, Samokhvalov'10, Iskin2007, Congjun} investigations of the FFLO state.

\begin{figure}
\includegraphics[width=3.4in,angle=0]{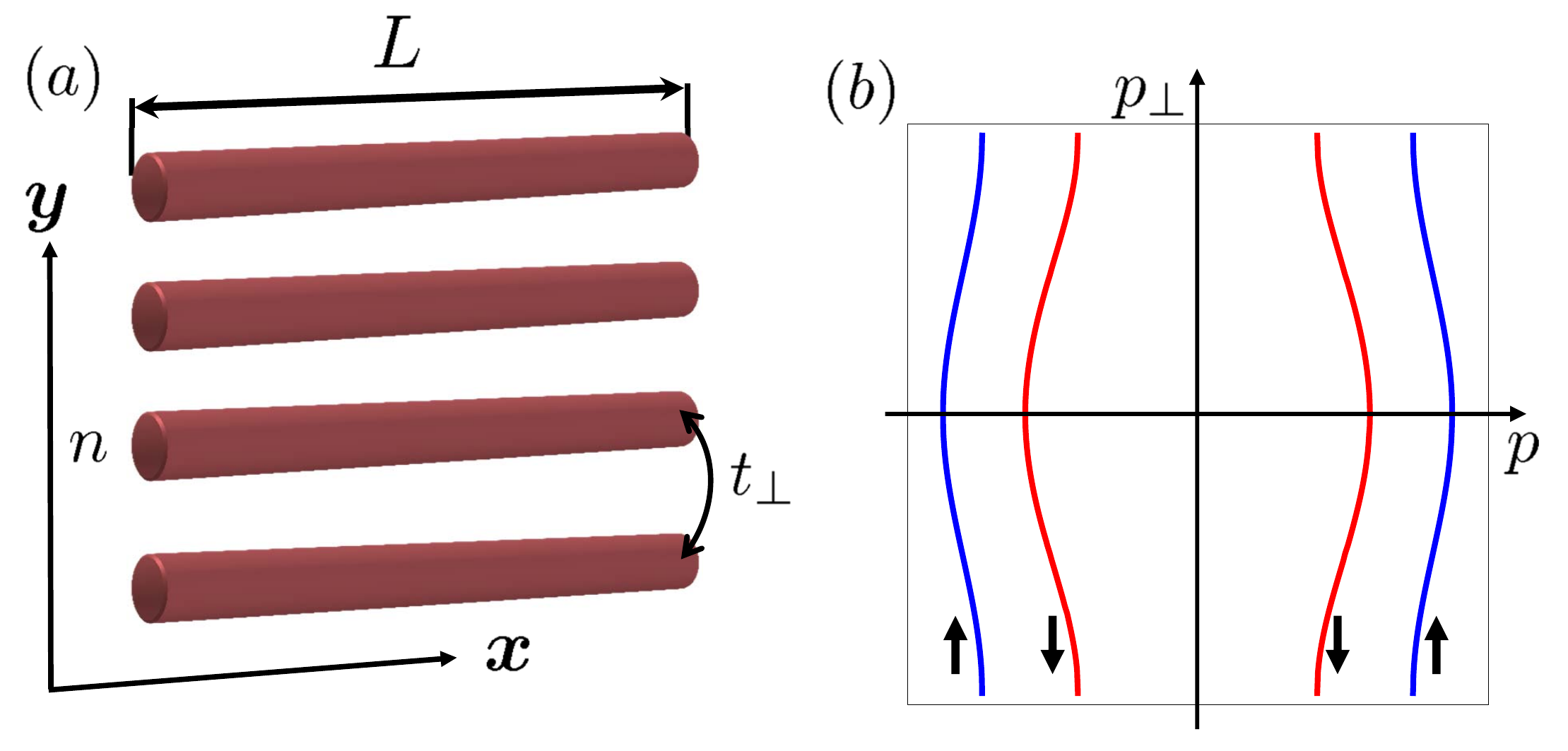}
\caption{(Color online) (a) An array of 1D tubes confining cold atoms, where $n$ is the tube number, and $t_{\perp}$ is the tunneling matrix element between adjacent tubes.  (b) Fermi surfaces for the majority ($\uparrow$) and minority ($\downarrow$) atoms in the quasi-1D limit $t_{\perp}\ll E_F$ (where $E_F$ is the Fermi energy of 1D motion along the tube).  The momentum components $p$ and $p_\perp$ are parallel and perpendicular to the tubes, respectively.}
\label{fig:3Dtubes}
\end{figure}

The FFLO state is expected to be more stable in one dimension (1D) than in three dimensions.  Thus, considerable experimental effort was made to realize trapping potentials in the form of an array of 1D tubes~\cite{Rice, Hulet2009}, as shown in Fig.~\ref{fig:3Dtubes}a.  Although such systems have a strong 1D anisotropy, it is important to realize that they are not strictly one-dimensional because of a non-zero tunneling amplitude $t_{\perp}$ between the tubes.  In the quasi-1D case $t_\perp\ll E_F$, where $E_F$ is the Fermi energy of 1D motion along the tubes, the system has an open, warped Fermi surface shown in Fig.~\ref{fig:3Dtubes}b.  Experimental results recently reported by the Rice group~\cite{Hulet2009} were obtained for $t_{\perp}\ll T$, where $T$ is the temperature.  This regime, essentially, corresponds to an incoherent mixture of 1D tubes, where each tube behaves independently.  If, however, the strength of the inter-tube tunneling is increased by lowering the confining optical-lattice potential in the transverse directions so that $t_\perp\gg T$, the system would cross over into the quasi-1D regime, where the superfluid phases on multiple tubes are locked together.  Such a system represents a quasi-1D fermionic superfluid.  Similar quasi-1D electronic systems have been studied extensively in solid-state physics, particularly among organic conductors and superconductors~\cite{Ishiguro-book, Lebed-book}. Quasi-1D superfluid states can be also realized in two and three-dimensional optical lattices using $p$-orbital bands~\cite{Congjun}.

Many theoretical papers studied strictly 1D fermionic systems with population imbalance using analytical approaches based on the Luttinger liquid~\cite{KunYang2001, Burovski2009} and the Bethe ansatz~\cite{Yang1967, Gaudin1967,  Lieb1968, Fuchs2004, Orso2007, Vincent_Liu, Luescher2008, Bolech2009}.  Numerical methods based on the density-matrix renormalization group and time-evolving block decimation~\cite{Feiguin2007, Feiguin2009, Heidrich-Meisner2010, Tezuka2008, Rizzi2008, Wang2009}, as well as quantum Monte Carlo~\cite{Batrouni2008, Casula2008}, were also employed.  Most of these approaches start from the 1D Luttinger liquid fixed point and take into account the inter-tube tunneling amplitude $t_{\perp}$ perturbatively.  This is justified as long as $t_\perp$ is much smaller than the superfluid transition temperature $T_c$, see, e.g.,\ Ref.~\cite{Yakovenko'85}.  However, in the opposite regime $t_\perp\gg T_c$ considered in our paper, the strictly 1D approaches are not applicable, because the quasiparticle dispersion substantially deviates from the 1D form.  In this domain, the quasi-1D system is more appropriately described within a Fermi-liquid picture and a mean-field theory of superfluid pairing.

The FFLO state in cold atoms has been recently studied in the quasi-1D geometry within a mean-field theory in Ref.~\cite{Parish2007}.  It has been known in the solid-state literature that, within a mean-field Bogoliubov-de Gennes (BdG) theory, an exact self-consistent inhomogeneous pairing potential for a quasi-1D system with population imbalance has the form of a soliton lattice.  The soliton-lattice solution was first obtained in the context of the Peierls model for charge-density waves in Refs.~\cite{Brazovskii1980, Horovitz'81, Mertsching1981,Brazovskii1984} (see reviews \cite{SovSciRev1984,Heeger1988}).  Subsequently, this model was mapped via a particle-hole transformation onto the FFLO pairing problem in quasi-1D superconductors \cite{Buzdin1983, Buzdin1987, Machida1984}.  Machida and Nakanishi \cite{Machida1984} applied these results to the superconducting material ErRh$_4$B$_4$ where a strong molecular ferromagnetic field is present.  Buzdin and Polonskii~\cite{Buzdin1987} studied a similar problem for organic superconductors, where imbalance between the spin-up and spin-down electrons can be induced by an external magnetic field.  An important result found in Refs.~\cite{Machida1984,Buzdin1987} is the existence of the second-order phase transition between the uniform and spatially-inhomogeneous superconducting states with an increase of the effective magnetic field $h$, which represents the difference between the chemical potentials of the spin-up and spin-down electrons.  When $h$ exceeds a critical value $h_c$, the superconducting order parameter develops a spatially-periodic modulation in the form of a soliton lattice.

In the present paper, we first summarize the self-consistent soliton-lattice solution of the BdG equations in the context of cold atoms and then focus on the observable physical properties of the soliton-lattice state.  In particular, we discuss three different spectroscopic methods for experimental detection and investigation of the soliton lattice in cold atoms. Optical spectroscopy of a soliton lattice for charge-density waves in conducting polymers was studied theoretically by Brazovskii and Matveenko~\cite{Matveenko'81}.  However, because of the difference in coherence factors between charge-density waves and superconductors, there results are not directly applicable to the quasi-1D fermionic superfluids.  Recently, there have been several numerical studies discussing properties of the FFLO state in 1D geometry~\cite{Luescher2008, Torma2008, Torma2010, Feiguin2009}.  However, these approaches are, strictly speaking, not applicable to the quasi-1D situation of our interest where $t_\perp\gg T_c$.  Signatures of the FFLO phase have been recently studied numerically for three-dimensional and quasi-1D optical lattices in Refs.~\cite{Loh'10,Trivedi'11}, which have some overlap with our results.

The paper is organized as follows.  In Sec.~\ref{sec:qualitative}, we qualitatively discuss a relationship between the FFLO state and the soliton lattice.  In Sec.~\ref{sec:exact}, we introduce the model Hamiltonian and review basic properties of the exact solution of the BdG equations for the quasi-1D case.  In Sec.~\ref{sec:experiment}, we propose and theoretically analyze several experiments for detection of the soliton lattice in cold-atom settings.  Finally, we conclude in Sec.~\ref{sec:discussion}. Technical details of mathematical derivations are relegated to Appendices.

\section{Qualitative discussion}\label{sec:qualitative}

\begin{figure}
\includegraphics[width=2.5in,angle=0]{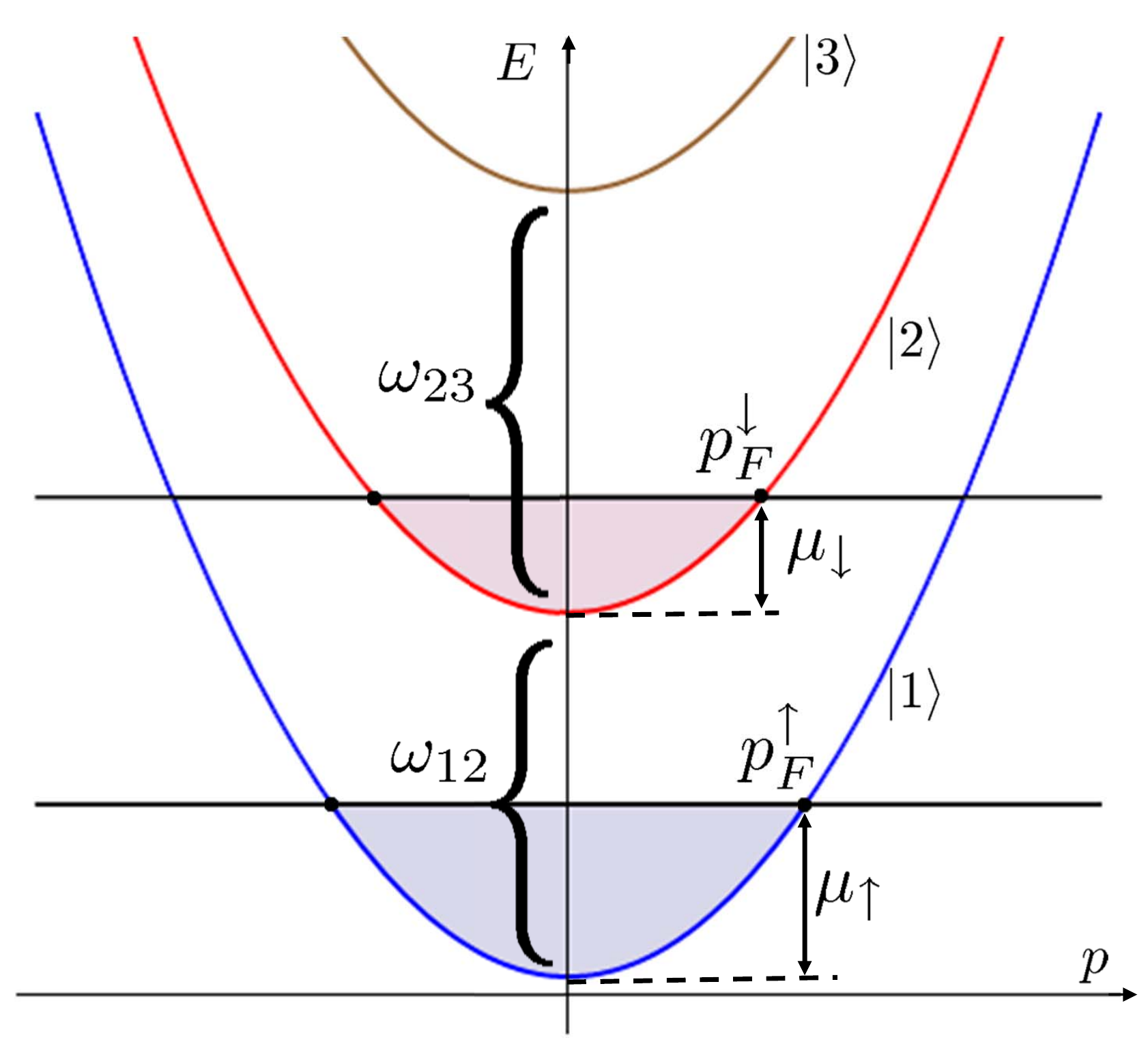}
\caption{(Color online) The 1D energy dispersion $E(p)$ of the fermionic atoms in the hyperfine states $|1\rangle$ and $|2\rangle$ with a population imbalance described by the chemical potentials $\mu_{\uparrow}$ and $\mu_{\downarrow}$ and the Fermi momenta $p_F^\uparrow$ and $p_F^\downarrow$.  The third, unpopulated state $|3\rangle$ can be used for detection purposes.  The atomic energy differences between the states $\ket{2}$ and $\ket{1}$ and the states $\ket{3}$ and $\ket{2}$ are denoted as $\omega_{12}$ and $\omega_{23}$, respectively.}
\label{fig:dispersion}
\end{figure}

Consider a fermionic atom, e.g.,\ $^6$Li, with two hyperfine states $\ket{1}$ and $\ket{2}$, e.g.,\ $F=\frac12$, $m_F=\frac12$ and $F=\frac12$, $m_F=-\frac12$, where $F$ and $m_F$ are the total spin and its projection along the quantization axis.  The energies of these two states differ by the Zeeman splitting $\omega_{12}$ due to an applied magnetic field, which will be not further discussed in this paper.  (The Planck constant $\hbar$ is set to unity everywhere.)  We denote these two states as the spin-up $\ket{1}\equiv\ket{\uparrow}$ and spin-down $\ket{2}\equiv\ket{\downarrow}$ states.  The atoms are loaded into the quasi-1D trap shown in Fig.~\ref{fig:3Dtubes}a, where they have a parabolic dispersion along the tubes and are confined in the transverse directions.  The energy dispersion of the states $\ket{1}$ and $\ket{2}$ is shown in Fig.~\ref{fig:dispersion} vs.\ the longitudinal momentum $p$ along the tubes for a fixed transverse momentum $p_\perp$.  By applying radio-frequency (rf) radiation with the frequency $\omega=\omega_{12}$, it is possible to induce transitions between the states $\ket{1}$ and $\ket{2}$ and, thus, control their relative populations.  These states are long-lived, so, after initialization of the system, the populations of the states $\ket{1}$ and $\ket{2}$ are fixed during the time of the experiment.  The states are characterized by the different chemical potentials $\mu_\uparrow$ and $\mu_\downarrow$, the 1D Fermi momenta $p_F^\uparrow$ and $p_F^\downarrow$, and the densities of atoms per unit length $\rho_\uparrow=p_\uparrow/\pi$ and $\rho_\downarrow=p_\downarrow/\pi$.  For concreteness, we assume that $\rho_\uparrow>\rho_\downarrow$ and refer to the $\ket{\uparrow}$ and $\ket{\downarrow}$ states as majority and minority.  The difference of the chemical potentials is $2h=\mu_\uparrow-\mu_\downarrow$, where $h$ is referred to as the effective magnetic field.  The difference of the Fermi momenta $Q=p_F^{\uparrow}-p_F^{\downarrow}$ results in the spin density per unit length $n_s=\rho_\uparrow-\rho_\downarrow=Q/\pi$. It is also useful to define the dimensionless spin polarization
\begin{align}\label{eq:polarization}
P=\frac{\rho_\uparrow-\rho_\downarrow}{\rho_\uparrow+\rho_\downarrow}.
\end{align}

\begin{figure}
\includegraphics[height=3in,angle=90]{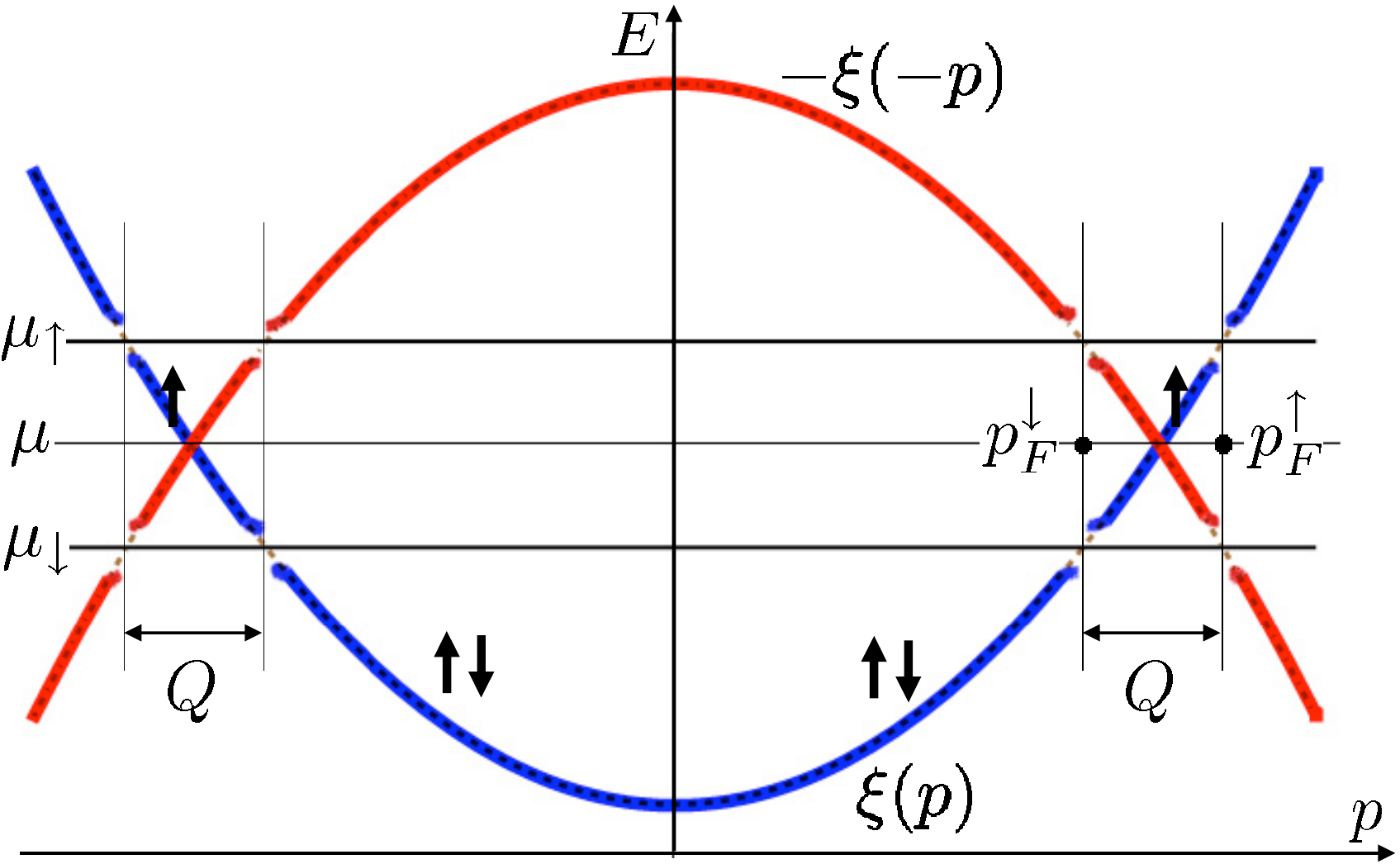}
\caption{(Color online)  Schematic plot of the quasiparticle energy dispersion $E(p)$ in the inhomogeneous superfluid phase.  Here $\xi(p)$ is the normal-state dispersion,  $2\mu=\mu_{\uparrow}+\mu_{\downarrow}$, and $Q=p_F^{\uparrow}-p_F^{\downarrow}$.}
\label{fig:phspectrum}
\end{figure}

An attractive $s$-wave interaction between the atoms results in the Cooper pairing of the fermionic atoms in the states $\ket{\uparrow}$ and
$\ket{\downarrow}$.  It is convenient to subtract the energy $\omega_{12}$ from the energy of the state $\ket{2}$ and, thus, align the bottoms of the two bands, as shown in Fig.~\ref{fig:phspectrum}.  In order to discuss the superconducting pairing, let us make the particle-hole transformation for the minority atoms, so that their dispersion relation becomes represented by the inverted parabola in Fig.~\ref{fig:phspectrum}.  Because of the mismatch of the Fermi momenta due to population imbalance, the conventional spatially-uniform BCS pairing potential is not favorable, since it would open an energy gap away from the chemical potentials of the atoms.  Larkin and Ovchinikov (LO)~\cite{LO} proposed that the pairing potential should be non-uniform and have the spatial dependence $\Delta(x)\propto\sin(Qx)$ \cite{Fulde-note}.  This order parameter couples fermions having the difference $\pm Q$ of the momenta $|p_\uparrow|$ and $|p_\downarrow|$, so it opens energy gaps at the chemical potentials for both majority and minority atoms, as shown in Fig.~\ref{fig:phspectrum}.  Then, the lower band is populated by the Bogoliubov quasiparticles with both spins up and down, the middle band is populated only by the quasiparticles with spin up, and the upper band is empty.

\begin{figure}
\includegraphics[width=3.4in,angle=0]{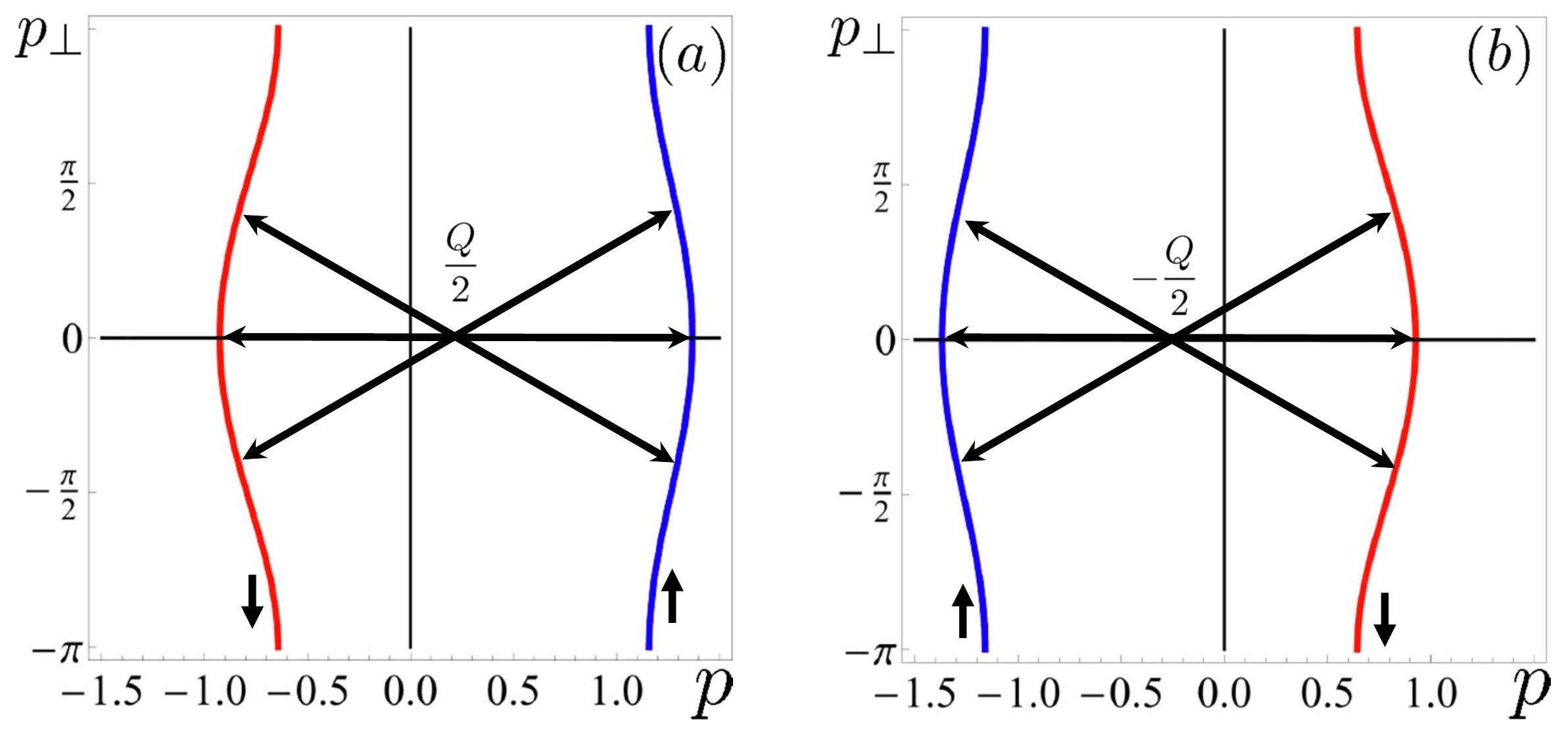}
\caption{(Color online) Schematic plot of the superfluid pairing for the LO order parameter.  In each panel, the centers of the three pairs of arrows, representing pairs of atoms on the Fermi surface, are located at the point with the momentum either $+Q/2$ or $-Q/2$, which is a half of the Cooper pair momentum in the LO phase.}
\label{fig:pairing}
\end{figure}

Fig.~\ref{fig:phspectrum} only illustrates the 1D dependence of the quasiparticle energy $E(p)$ on the longitudinal momentum $p$ in the LO state.  However, even for a non-zero $t_\perp$, the LO order parameter $\Delta(x)\propto\sin(Qx)$ still opens a gap on the whole quasi-1D Fermi surface, as illustrated in Fig.~\ref{fig:pairing} and explained below.  For simplicity, let us consider the tunneling amplitude $t_\perp$ between the tubes only in one transverse direction, as shown in Fig.~\ref{fig:3Dtubes}a, and denote the corresponding momentum $p_\perp$.  Then, the normal-state quasi-1D energy dispersion is
\begin{align}\label{eq:parabolic}
\epsilon_{\sigma}(p,p_\perp)=\frac{p^2}{2m}-2t_\perp\cos(p_\perp d)-\mu -\sigma h,
\end{align}
where $m$ is the mass of an atom, $d$ is the inter-tube spacing, and $\sigma h=\pm h$ for $\sigma=\uparrow,\downarrow$.  In the case where $t_\perp,\, h\ll\mu$, we can linearize the energy dispersion relation along the tube direction and obtain
\begin{align}\label{eq:linearized}
\epsilon_\sigma(p,p_\perp) \approx \pm v_F(p\mp p_F^\sigma)-2t_\perp\cos(p_\perp d),
\end{align}
where $v_F=p_F/m$ is the Fermi velocity, $p_F^\sigma=p_F+\sigma Q/2$, and $Q=2h/v_F$.  The Fermi surfaces $\epsilon_\sigma(p,p_\perp)=0$ obtained from Eq.~(\ref{eq:linearized}) are shown in Fig.~\ref{fig:3Dtubes}b.

As illustrated in Fig.~\ref{fig:pairing}, the majority and minority fermions on the $+p_F^\uparrow$ and $-p_F^\downarrow$ branches of the Fermi surface pair in such a manner that the total momentum of each pair is $+Q$ and does not dependent of the transverse momenta $\pm p_\perp$ of the paired fermions.  Correspondingly, the total momentum is $-Q$ for each pair on the $-p_F^\uparrow$ and $+p_F^\downarrow$ branches.  So, the LO order parameter with the two momentum components $\Delta(x)\propto(e^{iQx}-e^{-iQx})$ opens an energy gap on the whole Fermi surface for all values of $p_\perp$.  Thus, one expects to have a stable LO phase in the quasi-1D geometry.  However, this result is valid only for the linearized energy dispersion (\ref{eq:linearized}) and does not apply for a generic three-dimensional dispersion, e.g.,\ for a spherical Fermi surface.

The linearized dispersion (\ref{eq:linearized}) can be rewritten in the form
\begin{align}\label{eq:linearized+tilde}
\epsilon_\sigma \approx \pm v_F(p\mp \tilde p_F) -\sigma h,
\end{align}
where
\begin{align}\label{eq:pFtilde}
\tilde p_F(p_\perp) = p_F + \frac{2t_\perp}{v_F}\cos(p_\perp d)
\end{align}
is the longitudinal Fermi momentum as a function the transverse momentum $p_\perp$ at $h=0$.  As will be shown in Sec.~\ref{sec:exact}, it is possible to make a gauge transformation of the fermion operators and eliminate $\tilde p_F$ along with any explicit dependence on $p_\perp$ and $t_\perp$ from the Hamiltonian of the problem.  Physically, this gauge transformation corresponds to measuring the longitudinal momentum $p$ from the local Fermi momentum $\tilde p_F(p_\perp)$ at each point on the quasi-1D Fermi surface.  After this transformation, the mathematical problem formally becomes one-dimensional, and an exact solution of the mean-field equations can be obtained, which reduces to the LO order parameter and the soliton lattice in different limits.  Nevertheless, it is important to remember that, although $t_\perp$ can be formally eliminated from the mean-field Hamiltonian, the physical problem remains quasi-1D, and the presence of $t_\perp$ stabilizes fluctuations of the order parameter and makes the mean-field approach applicable.

Fig.~\ref{fig:phspectrum} illustrates the case where $h\gg\Delta$.  When $\Delta$ and $h$ are comparable, the two energy gaps in Fig.~\ref{fig:phspectrum} are coupled and cannot be treated independently.  The problem becomes mathematically complicated in this case.  Fortunately, an exact self-consistent solution of the BdG equation in the quasi-1D case can be obtained \cite{Brazovskii1980, Machida1984, Buzdin1987}.  The exact solution shows that, in general, $\Delta(x)$ is given by a periodic Jacobi elliptic function, such that it reduces to $\sin(Qx)$ in the limit $h\gg\Delta$.  The exact $\Delta(x)$ represents the so-called finite-zone potential, which opens only two gaps in Fig.~\ref{fig:phspectrum} at the chemical potentials $\mu_{\uparrow,\downarrow}$ and nowhere else.  The exact solution shows that there is a critical value $h_c=2\Delta_0/\pi$ such that the order parameter is uniform $\Delta(x)={\rm const}=\Delta_0$ for $h<h_c$, where $\Delta_0$ is the value of the BCS superconducting gap for $h=0$.

\begin{figure}
\includegraphics[width=3.4in,angle=0]{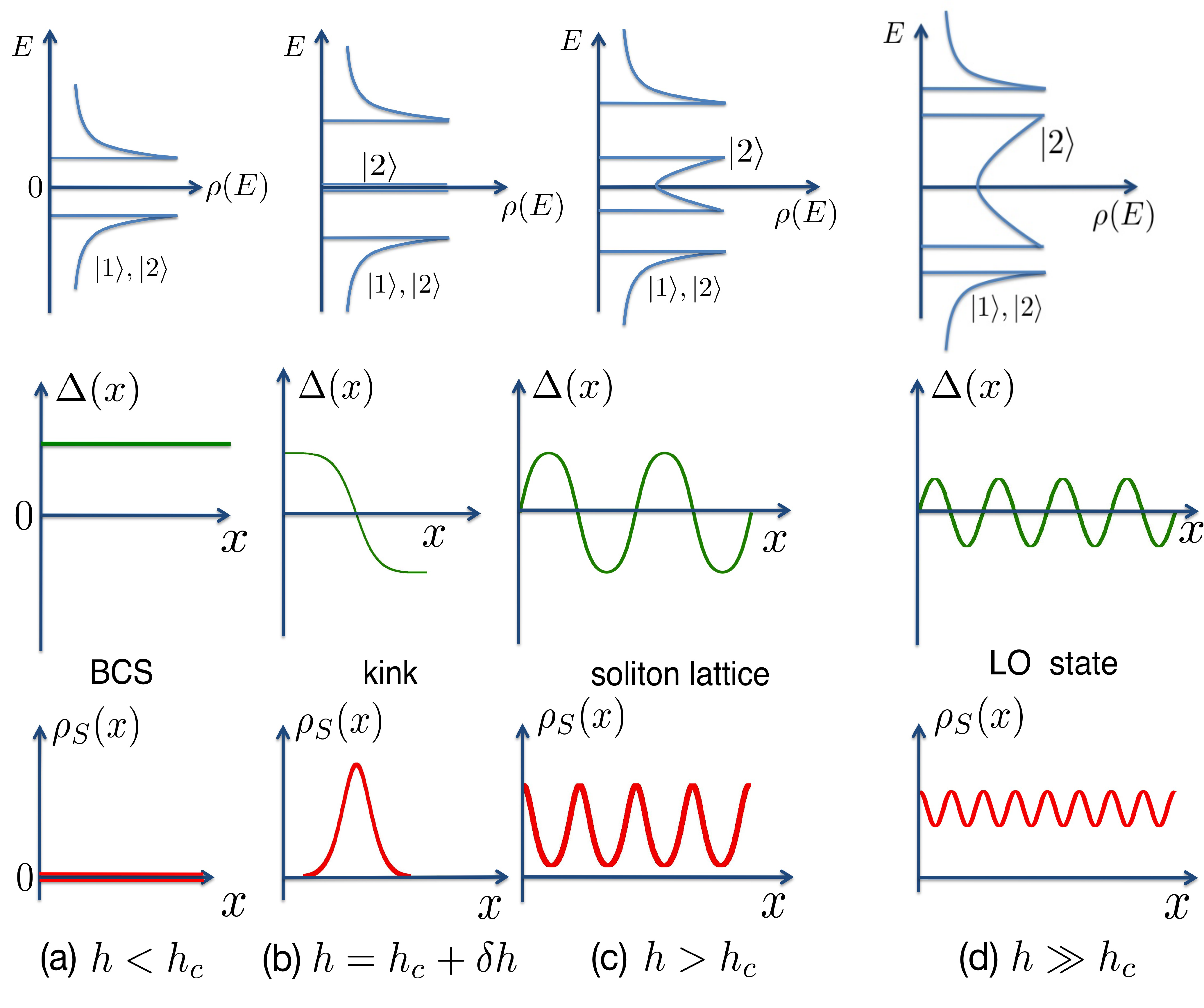}
\caption{(Color online) Schematic plots of the density of states (top row), superconducting order parameter $\Delta(x)$ (middle row), and the spin density $\rho_s(x)$ (bottom row) for different values of the effective magnetic field $h$.  (a) For $h<h_c$, the ground state of the system is uniform, as in the BCS theory.  (b) For one unpaired atom $h=h_c+\delta h$, $\Delta(x)$ has a kink soliton.  (c) For a small spin imbalance $h>h_c$, $\Delta(x)$ forms a soliton lattice.  (d) For a large imbalance $h\gg h_c$ corresponding to the LO limit, $\Delta(x)$ is sinusoidal.  Notice that $\rho_s(x)$ peaks at the points in space where $\Delta(x)$ vanishes.}
\label{fig:DoS}
\end{figure}

For $h$ slightly above $h_c$, the order parameter develops a series of kink solitons, where $\Delta(x)$ changes sign across each kink soliton.  In order to explain soliton formation qualitatively, let us consider the sequence of states shown in Fig.~\ref{fig:DoS}.  In this figure, the middle row shows the spatial dependence of $\Delta(x)$, the top row shows the density of quasiparticle states, and the bottom row shows the spatial dependence of the local spin density $\rho_s(x)$.  Panel (a) shows that, for $h<h_c$, the system has the conventional BCS order parameter with the uniform $\Delta(x)=\Delta_0$, a single gap in the density of states, and zero spin density $\rho_s=0$, i.e., no spin imbalance.  Now, suppose we add one extra majority atom to the system.  Because this atom cannot form a pair, it would have to populate an energy level above the gap, which would cost the energy $\Delta_0$.  However, if the system develops a kink, i.e.,\ a sign change of $\Delta(x)$, as shown in Panel (b), this configuration has an Andreev bound state in the middle of the gap, which can be occupied by the extra atom.  The total energy cost for creation of the soliton and occupation of the midgap state is $(2/\pi)\Delta_0$, which is lower than $\Delta_0$ for the uniform configuration.  Thus, the system spontaneous creates the kink soliton shown in Panel (b).  The spin density $\rho_s(x)$ of the extra atom is concentrated near the soliton.  Soliton creation has been discussed in the literature for charge-density waves~\cite{Brazovskii1980, Horovitz'81}, for mesoscopic superconducting wires~\cite{Yakovenko2002}, and in the cold-atom context~\cite{Yip'07}.

When a few atoms with spins up are added to the system, each atom creates a kink soliton.  The solitons repel each other and form a periodically spaced soliton lattice shown in Panel (c).  The order parameter $\Delta(x)$ experiences a series of sign changes.  The midgap states from different solitons hybridize and form a band in the middle of the main energy gap.  The spin density $\rho_s(x)$ consists of a series of spikes originating from each soliton.  When many atoms with spins up are added, $\Delta(x)$ becomes sinusoidal, corresponding to the LO limit, as shown in Panel (d).  The midgap band expands and occupies most of the former energy gap, leaving two small gaps above and below, in agreement with Fig.~\ref{fig:phspectrum}.  The spin density $\rho_s(x)$ is sinusoidally modulated with a small amplitude and the wavelength a half of that for $\Delta(x)$.

In the next Section, we present a detailed mathematical derivation of the results qualitatively discussed above.  Then, in Sec.~\ref{sec:experiment}, we discuss three spectroscopic techniques for experimental detection of the soliton lattice.

\section{Theoretical model and exact solution}\label{sec:exact}

Let us consider a two-dimensional (2D) array of parallel 1D tubes in the $x$ direction with the inter-tube spacings $d_y$ and $d_z$ in the $y$ and $z$ directions, as sketched in Fig.~\ref{fig:3Dtubes}a.  In the second-quantized formalism, the single-particle Hamiltonian for this quasi-1D system is
\begin{align}\label{Eq1}
\hat H_0&=\sum\limits_{\bm n,\sigma}\int dx \, \hat\psi_{\bm n,\sigma}^\dagger(x)
\left(-\frac{\partial_x^2}{2m} - \mu_\sigma \right) \hat\psi_{\bm n,\sigma}(x) \nonumber\\
&+ \sum\limits_{\langle\bm n,\bm n'\rangle,\sigma}\int dx\, t_\perp
\left[\hat\psi_{\bm n,\sigma}^\dagger(x) \hat\psi_{\bm n',\sigma}(x) + \rm H.c.\right],
\end{align}
where $\psi_{\bm n,\sigma}(x)$ is the fermion annihilation operator for an atom in the state $\sigma=\uparrow,\downarrow$ on the tube $\bm n$, where $\bm n=(n_y,n_z)$ is the 2D index of the tube, and $t_\perp$ the amplitude of tunneling between adjacent tubes.  An attractive interaction between the atoms leads to the BCS pairing, which, at the mean-field level, is described by the Hamiltonian
\begin{align}\label{Hi}
\hat H_{\rm SC}\!=\!-&\!\sum_{\bm n}\int\! dx\! \left[\Delta_{\bm n}(x)
\hat\psi_{\bm n,\up}^\dagger(x) \hat\psi_{\bm n,\dn}^\dagger(x) + \rm H.c.\right],
\end{align}
where the pairing potential is determined self-consistently
\begin{align}\label{eq:SCF}
\Delta_{\bm n}(x)=g\,\langle\hat\psi_{\bm n,\dn}(x)\hat\psi_{\bm n,\up}(x)\rangle
\end{align}
with the $s$-wave interaction amplitude $g$.  We assume that pairing is local in the real space, and the inter-tube pairing potential can be neglected.  Given that we consider identical tubes, we take the pairing potential to be independent of $\bm n$, i.e., $\Delta_{\bm n}(x)=\Delta(x)$.  We study the problem at zero temperature $T=0$, so the brackets in Eq.~\eqref{eq:SCF} represent averaging with respect to the ground state.

After the Fourier transform $\psi_\sigma(x,\bm p_\perp)=\sum_{\bm n}
e^{-i\bm p_\perp\cdot\bm\rho_{\bm n}}\psi_{\bm n,\sigma}(x)$, where $\bm p_\perp=(p_y,p_z)$ is the
transverse momentum, and $\bm\rho_{\bm n}=(n_yd_y,n_zd_z)$ is the 2D vector in the $(y,z)$ plane, the full Hamiltonian $\hat H=\hat H_0+\hat H_{\rm SC}$ becomes
\begin{align}
\hat H &= \! \sum\limits_{\bm p_\perp,\sigma} \int\! dx \,
\hat\psi_\sigma^\dagger(x,\bm p_\perp)\!
\left[-\frac{\partial_x^2}{2m}\!+\!\xi(\bm p_\perp)\!-\!\mu\!-\!\sigma h \right] \!
\hat\psi_\sigma(x,\bm p_\perp) \nonumber\\
&+\sum_{\bm p_\perp} \int dx \left[\Delta(x) \,
\hat\psi_\up^\dagger(x,\bm p_\perp) \hat\psi_\dn^\dagger(x,-\bm p_\perp)+ \rm H.c.\right].
\label{eq:H}
\end{align}
Here $\xi(\bm p_\perp)=-2t_\perp[\cos(p_yd_y)+\cos(p_zd_z)]$ is the transverse dispersion, and
the sum over the transverse momenta means $\sum_{\bm p_\perp}=d_yd_z\int_0^{2\pi/d_y}dp_y\int_0^{2\pi/d_z}dp_z/(2\pi)^2$.
The Hamiltonian (\ref{eq:H}) is the sum $\hat H=\sum_{\bm p_\perp}\hat H_{\rm 1D}(\bm p_\perp)$ of 1D Hamiltonians $H_{\rm 1D}$ with fixed values of $\bm p_\perp$
\begin{align}\label{H1D}
\hat H_{\rm 1D} & = \sum_\sigma\int dx \, \hat\psi_\sigma^\dagger(x)
\left[-\frac{\partial_x^2}{2m} - \tilde\mu - \sigma h \right] \hat\psi_\sigma(x) \nonumber\\
& + \int dx \left[\Delta(x)\,\hat\psi_\up^\dagger(x)\,\hat\psi_\dn^\dagger(x)+ \rm H.c.\right],
\end{align}
where $\tilde\mu(\bm p_\perp)=\mu -\xi(\bm p_\perp)$ is the renormalized chemical potential, and the argument $\bm p_\perp$ in $\hat\psi_\sigma(x)$ is implied.  The self-consistency condition (\ref {eq:SCF}) has the form
\begin{align}\label{eq:self}
\Delta(x)=g\,\sum_{\bm p_\perp}\langle\hat\psi_\dn(x,\bm p_\perp)\,
\hat\psi_\up(x,-\bm p_\perp)\rangle.
\end{align}

The integrand in the Hamiltonian (\ref{H1D}) can be written in the matrix form as
\begin{align}\label{H-matrix}
\left[\begin{matrix}
\hat\psi_\uparrow^\dagger(x) \\ \hat\psi_\downarrow(x)
\end{matrix} \right]^T
\begin{pmatrix}
-\frac{\partial_x^2}{2m}-\tilde\mu-h \!&\ \Delta(x)\\
\overline\Delta(x)  \!&\  \frac{\partial_x^2}{2m}+\tilde\mu-h
\end{pmatrix}
\left[\begin{matrix}
\hat\psi_\uparrow(x) \\ \hat\psi_\downarrow^\dagger(x)
\end{matrix} \right],
\end{align}
where the overline in $\overline\Delta$ denotes complex conjugation.
The Hamiltonian $\hat H_{\rm 1D}$ also acquires a constant term $\sum_p(p^2/2m-\tilde\mu+h)$ originating from the fermion commutation relation
$\hat\psi_\downarrow^\dagger\hat\psi_\downarrow=1-
\hat\psi_\downarrow\hat\psi_\downarrow^\dagger$.  The matrix (\ref{H-matrix}) can be diagonalized by the Bogoliubov transformation
\begin{align}\label{Bogtrans}
\begin{split}
\hat\psi_\uparrow(x)&=\sum\limits_\lambda
\left[U_\lambda(x) \, \hat\gamma_{\lambda,\uparrow}-
\overline V_\lambda(x) \, \hat\gamma_{\lambda,\downarrow}\dg\right],\\
\hat\psi_\downarrow^\dagger(x)&=\sum\limits_\lambda
\left[\overline U_\lambda(x) \, \hat\gamma_{\lambda,\downarrow}\dg+
V_\lambda(x) \, \hat\gamma_{\lambda,\uparrow}\right],
\end{split}
\end{align}
where the sums are taken over the eigenstates $\lambda$ specified below.
The coefficients $U_\lambda(x)$ and $V_\lambda(x)$ are selected so that the Nambu spinor $\Psi_\lambda(x)=[U_\lambda(x),V_\lambda(x)]^T$ is an eigenstate of the BdG equation with an eigenvalue $E_\lambda$
\begin{align}\label{BdG3D}
\begin{pmatrix}
-\frac{\partial_x^2}{2m}-\tilde\mu & \Delta(x)\\
\overline\Delta(x)  &  \frac{\partial_x^2}{2m}+\tilde\mu
\end{pmatrix} \Psi_\lambda(x)
=E_\lambda\Psi_\lambda(x).
\end{align}
We also impose the normalization condition
\begin{align}
\int_{-L}^{L} dx\left[|U_\lambda(x)|^2+|V_\lambda(x)|^2\right]=1,
\label{eq:norm}
\end{align}
where $2L$ is the length of the system.  The BdG equation~(\ref{BdG3D}) has the particle-hole symmetry: if $\left[U_\lambda(x), V_\lambda(x)\right]$ is an eigenstate with the energy $E_\lambda$, then
$\left[-\overline V_\lambda(x),\overline U_\lambda(x),\right]$ is also an eigenstate with the energy $-E_\lambda$:
\beg\label{reflect}
\left[\begin{matrix} U_\lambda(x) \\ V_\lambda(x) \end{matrix} \right]\to
\left[\begin{matrix} -\overline V_\lambda(x) \\ \overline U_\lambda(x) \end{matrix}\right], \quad E_\lambda\to -E_\lambda.
\en
This property allows us to restrict summations over eigenenergies in subsequent calculations only to the positive values $E_\lambda>0$.

Substituting Eq.~\eqref{Bogtrans} into Eq.~\eqref{H-matrix} and utilizing the properties of $U_\lambda(x)$ and $V_\lambda(x)$ outlined above, we diagonalize the Hamiltonian $\hat H_{\rm 1D}$
\begin{align}\label{H-diagonal}
\hat H_{\rm 1D}& = \sum_{E_\lambda>0}\sum\limits_{\sigma=\uparrow,\downarrow}(E_\lambda-\sigma h)\,\hat\gamma^\dag_{\lambda,\sigma}\hat\gamma_{\lambda,\sigma}
+ {\cal E}_0.
\end{align}
Here ${\cal E}_0$ is the reference energy given by \cite{Machida1984,Yakovenko2002}
\begin{align}\label{eq:E0}
{\cal E}_0=-\sum_{E_\lambda>0}E_\lambda+\sum_p\xi_p,
\end{align}
where the first sum originates from the fermion commutation relation
$\hat\gamma_{\lambda,\downarrow}\hat\gamma_{\lambda,\downarrow}^\dag=1-
\hat\gamma_{\lambda,\downarrow}^\dag\hat\gamma_{\lambda,\downarrow}$, and
\begin{align}\label{xi_p}
\xi_p=\frac{p^2}{2m}-\tilde\mu
\end{align}
is the normal-state dispersion relation.
The occupation numbers of the single-particle states for the Hamiltonian~\eqref{H-diagonal} are
\begin{align} \label{vacuum}
&\langle\gamma^\dag_{\lambda,\sigma}\gamma_{\lambda,\sigma}\rangle
=n_F(E_\lambda-\sigma h),
\end{align}
where $n_F(E)$ is the Fermi distribution function, which reduces to the step function $n_F(E)=\theta(-E)$ at $T=0$.  The ground state energy of the system is \cite{Machida1984,Yakovenko2002}
\begin{align}\label{eq:F0}
{\cal F}_0={\cal E}_0+\sum_{E_\lambda>0,\sigma}(E_\lambda-\sigma h)n_F(E_\lambda-\sigma h)+\int\limits_{-L}^L \frac{\Delta^2(x)}{|g|}dx,
\end{align}
where the last term originates from the Hubbard-Stratonovich transformation of the interaction between the fermions.

The pairing potential $\Delta(x)$ in Eq.~\eqref{BdG3D} can be selected to be real.  Now we linearize the dispersion relation in the longitudinal direction near the Fermi surface and use the approximation similar to Eq.~\eqref{eq:linearized+tilde} in the BdG equation
\begin{align}\label{BdG3D1}
&\begin{pmatrix}
v_F[i(-1)^\alpha\partial_x - \tilde p_F] \!&\ \Delta(x) \\
\Delta(x) \!&\  -v_F[i(-1)^\alpha\partial_x - \tilde p_F]
\end{pmatrix}\!\Psi_\lambda^{(\alpha)}(x)\!\nonumber\\
&=E_\lambda\Psi_\lambda^{(\alpha)}(x),
\end{align}
where
\begin{align}
\label{eq:pFtilde-again}
\tilde p_F=\sqrt{2m\tilde \mu}\approx p_F-\xi(\bm p_\perp)/v_F
\end{align}
is the renormalized Fermi momentum, and the index $\alpha=1,2$ corresponds to the right- and left-moving atoms.  We seek solutions of the BdG equation in the form $\Psi_\lambda(x)=\sum_\alpha\Psi_\lambda^{(\alpha)}(x)$ with
\begin{align}
\Psi^{(\alpha=1)}_{\lambda}(x)&\equiv\left(\begin{matrix}
U_{\lambda,1}(x) \\ V_{\lambda,1}(x)\end{matrix}\right)=\left(
\begin{matrix} u_\lambda(x) \\ v_\lambda(x) \end{matrix}\right)e^{i\tilde{p}_Fx},  \nonumber\\
\label{eq:UnVn}\\
\Psi^{(\alpha=2)}_{\lambda}(x)&\equiv\left(\begin{matrix}
U_{\lambda,2}(x) \\ V_{\lambda,2}(x)\end{matrix}\right)=\left(
\begin{matrix} v_\lambda(x) \\ u_\lambda(x) \end{matrix}\right)e^{-i\tilde{p}_Fx}.\nonumber
\end{align}
In Eq.~\eqref{eq:UnVn}, we used the symmetry between the $\alpha=1$ and $2$ components: $u_\lambda\equiv u_{\lambda,1}=v_{\lambda,2}$ and $v_\lambda\equiv v_{\lambda,1}=u_{\lambda,2}$.
Substituting Eq.~\eqref{eq:UnVn} into Eq.~\eqref{BdG3D1}, we eliminate $\tilde p_F(\bm p_\perp)$ from the BdG equation and obtain the 1D equation for the slowly varying envelope functions $u_\lambda(x)$ and  $v_\lambda(x)$
\beg\label{BdG1D}
\left(\begin{matrix} -iv_F\partial_x & \Delta(x) \\ \Delta(x) & iv_F\partial_x \end{matrix}
\right)
\left(\begin{matrix} u_\lambda(x) \\ v_\lambda(x) \end{matrix}\right)
= E_\lambda\left(\begin{matrix} u_\lambda(x) \\ v_\lambda(x) \end{matrix}
\right).
\en
In terms of the amplitudes $u_\lambda$ and $v_\lambda$, the self-consistency condition~(\ref{eq:self}) now reads
\beg\label{SCC}
\begin{split}
\Delta(x)={g}\sum\limits_{\lambda}&[u_\lambda(x)\overline{v}_\lambda(x)+\overline{u}_\lambda(x)v_\lambda(x)]\\&\times\left[1-n_F(E_\lambda+h)-n_F(E_\lambda-h)\right].
\end{split}
\en
Summation over $\bm p_\perp$ in Eq.~(\ref{SCC}) is omitted, because solutions of Eq.~\eqref{BdG1D} do not depend on $\bm p_\perp$, and the phase factors from Eq.~\eqref{eq:UnVn} containing $\tilde p_F(\bm p_\perp)$ cancel out in Eq.~(\ref{SCC}).  Thus, as a result of the linearization approximation, we managed to eliminate the transverse attributes $t_\perp$ and $\bm p_\perp$ from the BdG equation \eqref{BdG1D} and the self-consistency condition \eqref{SCC} and reduce the ground-state mean-field problem to a purely 1D formulation, effectively corresponding to a single tube.  Deviations of the actual dispersion relation  from the linearization approximation are of the order of $t_\perp^2/E_F$, so the approximation employed in this paper is applicable when $t_\perp^2/E_F\ll T_c \ll t_\perp\ll E_F$.

Although $t_\perp$ has been eliminated from calculation of the mean-field ground state, a non-zero value of $t_\perp$ is crucially important when considering fluctuations of the order parameter near the ground-state configuration.  While the ground state $\Delta(x)$ depends only on the coordinate $x$, the fluctuating order parameter depends on all three coordinates $\Delta(x,y,z)$.  A non-zero tunneling amplitude $t_\perp$ produces transverse phase stiffness in the effective action, which is proportional to $t_\perp^2[(\partial_y \Phi)^2+(\partial_z \Phi)^2]$ with $\Phi$ being the phase of the superconducting order parameter.  As a result, phase fluctuations are suppressed due to the three-dimensional anisotropic stiffness (as opposed to 1D stiffness for uncoupled chains), so the true long-range order is stabilized at low enough temperatures, and the mean-field approximation is justified.  Stability of the FFLO phase in quasi-1D geometry was recently confirmed numerically in Ref.~\cite{Trivedi'11}.

A self-consistent solution of Eqs.~\eqref{BdG1D} and \eqref{SCC} was first derived in Refs.~\cite{Brazovskii1980, Horovitz'81} in the context of charge-density waves and than subsequently extended to various other systems~\cite{Mertsching1981,Brazovskii1984}, including inhomogeneous superconductors~\cite{Buzdin1983, Machida1984, Buzdin1987}.  To keep our discussion self-contained, we outline the properties of the solution and refer the reader to the literature for further details.  Let us introduce the functions
\beg\label{fuv}
f_\lambda^{\pm}=\frac{1}{\sqrt{2}}(u_\lambda\pm iv_\lambda),
\en
and rewrite Eq.~\eqref{BdG1D} in the following form:
\begin{align} \label{eq:AA}
\left(\begin{array}{cc}
  0 & \hat A \\
  \hat A^\dag & 0
\end{array}\right) \left(\begin{array}{c}
                     f_{\lambda}^{+} \\
                     f_{\lambda}^{-}
                   \end{array}\right)=E_{\lambda}\left(\begin{array}{c}
                     f_{\lambda}^{+} \\
                     f_{\lambda}^{-}
                   \end{array}\right),
\end{align}
where $\hat A=i[-v_F \partial_x+\Delta(x)]$ and $\hat A^\dag=-i[v_F \partial_x+\Delta(x)]$ are mutually adjoint operators.  After simple manipulations, Eq.~\eqref{eq:AA} can be written in the supersymmetric (SUSY) form
\begin{align}\label{eq:SUSY}
\left(\begin{array}{cc}
  \hat A \hat A^\dag & 0 \\
  0 & \hat A^\dag \hat A
\end{array}\right) \left(\begin{array}{c}
                     f_{\lambda}^{+} \\
                     f_{\lambda}^{-}
                   \end{array}\right)=E^2_{\lambda}\left(\begin{array}{c}
                     f_{\lambda}^{+} \\
                     f_{\lambda}^{-}
                   \end{array}\right).
\end{align}
Equation~\eqref{eq:SUSY} corresponds to the $N=2$ SUSY quantum mechanics introduced by Witten~\cite{Junker_book}, where $\Delta(x)$ is the SUSY potential. This connection will be important for a discussion of zero-energy bound states at domain walls.  In the explicit form, Eq.~(\ref{eq:SUSY}) reads
\beg\label{eqfs}
\begin{split}
&\left(v_F^2\frac{d^2}{dx^2}+E_\lambda^2-\Delta^2(x)
\pm v_F\frac{d\Delta(x)}{dx}\right)f_\lambda^{\pm}(x)=0,
\end{split}
\en
while the self-consistency equation (\ref{SCC}) is
\beg\label{SCCf}
\begin{split}
\Delta(x)=g\sum\limits_{\lambda}&\frac{1}{2E_\lambda}\left[v_F\partial_x+2\Delta(x)\right]|f_{\lambda}^{+}(x)|^2\\&\times\left[1-n_F(E_\lambda+h)-n_F(E_\lambda-h)\right].
\end{split}
\en
The Schr\"{o}dinger equations (\ref{eqfs}), with the effective potentials determined by the order parameter $\Delta(x)$, are integrable for a special type of reflectionless potentials.  A solution for $\Delta(x)$ minimizing the ground-state energy ${\cal F}_0$ can be written in terms of the Jacobi elliptic functions specified by the modulus $k$
\beg\label{iden}
\frac{\Delta(x)}{\Delta_2}\!=\!(1\!-\!k')\text{sn}[(1+k')\zeta,\nu]=k^2\frac{\text{sn}(\zeta,k)\text{cn}(\zeta,k)}{\text{dn}(\zeta,k)},
\en
where
\beg\label{eq:zeta}
\zeta=x\,\frac{\Delta_2}{v_F}, \qquad k'=\sqrt{1-k^2}, \qquad \nu=\frac{1-k'}{1+k'}.
\en
The second equality in Eq.~\eqref{iden} follows from the properties of the elliptic functions.  The parameter $\Delta_2$ and the modulus $k$ of the Jacobi elliptic function have to be determined from the minimization of the ground-state energy ${\cal F}_0$ in Eq.~\eqref{eq:F0}, see Ref.~\cite{Machida1984} for more details.  The period $l$ of the order parameter $\Delta(x)$ is
\beg\label{eq:l}
l=\frac{2\pi}{Q}=\frac{2v_F}{\Delta_2}{\mathrm K}(k),
\en
where ${\mathrm K}(k)$ is the complete elliptic integral of the first kind.  Upon substitution of Eq.~\eqref{iden} into the ground-state energy ${\cal F}_0$ and minimization with respect to $\Delta_2$ and $k$, one finds~\cite{Brazovskii1980, Machida1984}
\beg\label{param}
\frac{\Delta_0}{\Delta_2}=k, \quad \frac{2{\mathrm E}(k)}{\pi}\Delta_2=h,
\en
where $\Delta_0$ is the BCS energy gap for a homogeneous unpolarized system at $h=0$, and ${\mathrm E}(k)$ is the complete elliptic integral of the second kind.  The parameter $\Delta_0$ is a convenient way to characterize the strength of the attractive interaction $g$ in the system.  Eliminating $\Delta_2$ from Eq.~\eqref{param}, we obtain an equation for the modulus $k$ in terms of the given values of the effective magnetic field $h$ and the BCS gap $\Delta_0$:
\begin{align}\label{eq:k}
\frac{k}{{\mathrm E}(k)}=\frac{2}{\pi}\frac{\Delta_0}{h}.
\end{align}
The values of $k$ are restricted to $k\leq1$ because of the properties of the elliptic functions.  With the increase of $h$, Eq.~\eqref{eq:k} acquires a solution at the critical value $h_c=2\Delta_0/\pi$, where $k\to 1$ and ${\mathrm E}(k=1)=1$.  At $h=h_c$, the solution $\Delta(x)$ corresponds to a single soliton, as shown in Panel (b) in Fig.~\ref{fig:DoS}.  With the further increase of $h$, the value of $k$ given by Eq.~\eqref{eq:k} decreases, so the soliton lattice period $l$ in Eq.~\eqref{eq:l} also decreases, corresponding to Panel (c) in Fig.~\ref{fig:DoS}.  In the LO limit $h\gg\Delta_0$, Eq.~\eqref{eq:k} gives $k\ll1$, and the period $l$ in Eq.~\eqref{eq:k} becomes $l=\pi v_F/h$.  In this limit, the order parameter takes the LO form
$\Delta(x)\approx(\Delta_2 k^2/2)\sin(2\pi x/l)$.

\begin{figure}
\includegraphics[height=2.7in,angle=0]{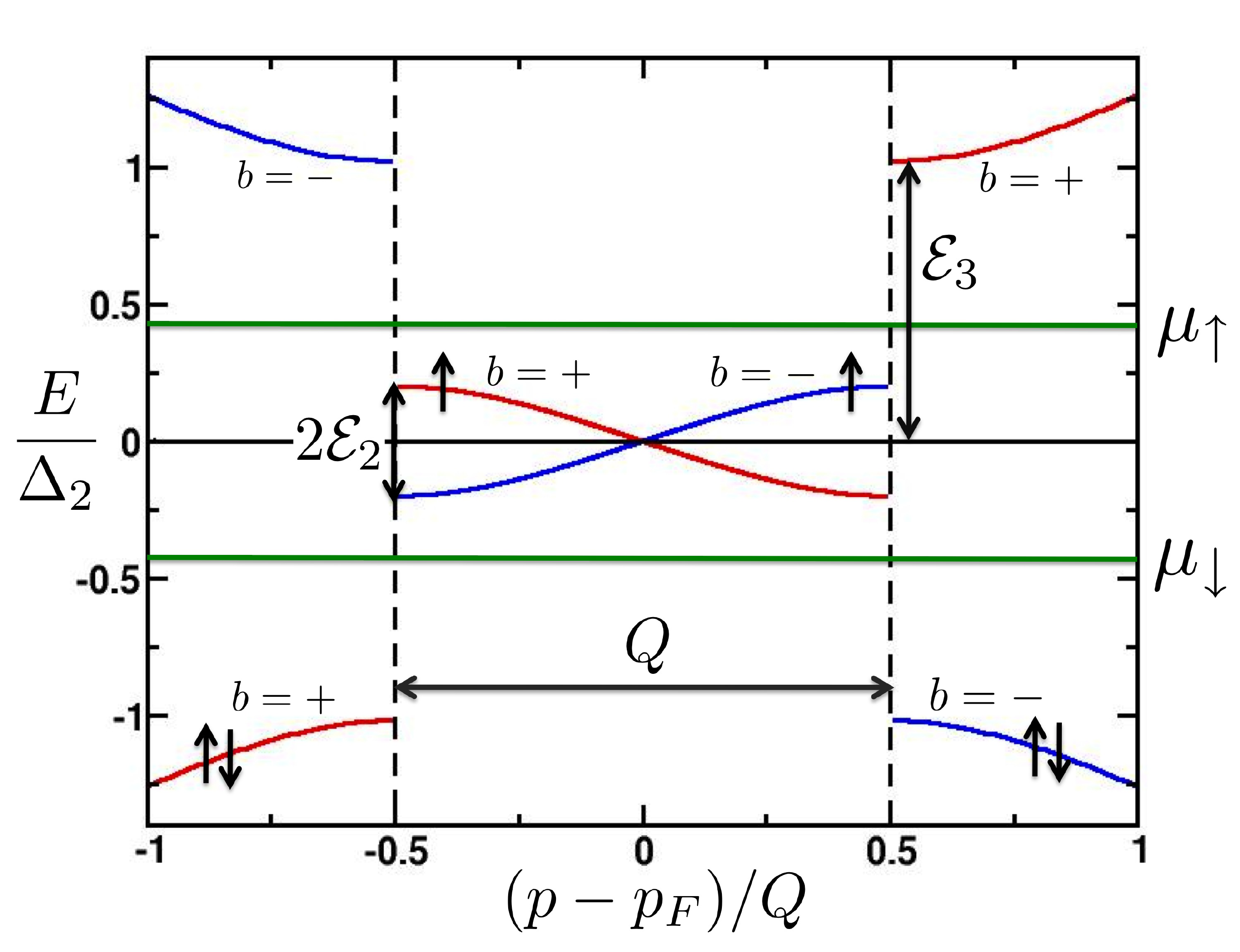}
\caption{(Color online) The single-particle energy dispersion relation $E(p)$ in the quasi-1D FFLO state described by Eq.~(\ref{iden}).  The spectrum is obtained by solving Eqs.~(\ref{q2E2}) and (\ref{q2E3}) for given values of the momentum $p$ and the parameter $b=\pm$.  This plot is a zoom into Fig.~\ref{fig:phspectrum} near the Fermi momentum $p_F$.}
\label{fig:exactdispersion}
\end{figure}

The solution for the amplitudes $f^\pm(x)$ can be expressed in terms of the function $\gamma(\zeta)$ satisfying the following equation
\beg\label{eq:E23}
\Delta_2^2\left(\frac{d\gamma}{d\zeta}\right)^2=4\gamma(\gamma-{\cal E}_2^2)({\cal E}_3^2-\gamma),
\en
where the variable $\zeta$ is defined in Eq.~\eqref{eq:zeta}.
Here we introduced the parameters ${\cal E}_3=\Delta_2$ and ${\cal E}_2=\Delta_2k'$.  The parameters ${\cal E}_2$ and ${\cal E}_3$ are the band edges in the single-particle excitation spectrum shown in Fig.~\ref{fig:exactdispersion}.  The spectrum in Fig.~\ref{fig:exactdispersion} corresponds to the spatially-inhomogenous order parameter $\Delta(x)$ in Eq.~(\ref{iden}) and represents a zoom-in of the spectrum shown in Fig.~\ref{fig:phspectrum} in the vicinity of the Fermi momentum $p\approx p_F$.

The solution of Eq.~\eqref{eq:E23} has the form
\beg
2\gamma(\zeta)={\cal E}_2^2+{\cal E}_3^2-\Delta^2(\zeta)+\Delta_2\frac{d\Delta(\zeta)}{d\zeta}.
\en
One can check by direct substitution that the solution of Eq.~(\ref{eqfs}) reads
\beg\label{fspm}
\begin{split}
f_{\lambda,b}^+(x)&=
\sqrt{\frac{E_\lambda^2-\gamma(x)}{2LA_\lambda}}\exp\left[\frac{ib}{v_F}\int\limits_{0}^x\frac{\sqrt{R_\lambda}dx'}{E_\lambda^2-\gamma(x')}\right],
\end{split}
\en
where $R_\lambda=E_\lambda^2(E_\lambda^2-{\cal E}_2^2)(E_\lambda^2-{\cal E}_3^2)$, $2L$ is the length of the system, and $A_\lambda$ is a normalization factor \cite{Brazovskii1984}
\beg\label{Alam}
A_\lambda=E_\lambda^2-{\cal E}_3^2\frac{{\mathrm E}(k)}{{\mathrm K}(k)}.
\en
The index $b=\pm$ in Eq.~(\ref{fspm}) distinguishes quasiparticles residing on the two energy branches of the spectrum shown in Fig.~\ref{fig:exactdispersion} and allows us to define unambiguously the momentum $p_{\lambda,b}$ for a given value of $E_\lambda$.  Using Eqs.~\eqref{BdG1D} and \eqref{fuv}, we find the other solution
\begin{align}\label{fspm1}
f_{\lambda,b}^-(x) = c_{\lambda,b}\,e^{-i\varphi_{\lambda,b}}f_{\lambda,b}^+(x+l/2),
\end{align}
where the coefficient $c_{\lambda,b}$ and the phase $\varphi_{\lambda,b}$ are
\begin{equation}
\begin{split}
c_{\lambda,b}&=\textrm{sign}(bE_\lambda) \times
\left\{\begin{matrix} -1, \quad |E_\lambda|<{\cal E}_2, \\
+1, \quad |E_\lambda|>{\cal E}_3,
\end{matrix}
\right. \\
\varphi_{\lambda,b}&=\frac{b}{v_F}\int\limits_{0}^{l/2}\frac{\sqrt{R_\lambda}dy}{E_\lambda^2-\gamma(y)}.
\end{split}
\end{equation}
We emphasize that, in order to recover correct amplitudes $u_{\lambda,b}$ and $v_{\lambda,b}$, it is necessary to know the relative phase between $f_{\lambda,b}^+(x)$ and $f_{\lambda,b}^-(x)$ in Eq.~\eqref{fspm1}.  This subject was not discussed in the previous work focusing on thermodynamics of the soliton-lattice state~\cite{Brazovskii1980, Mertsching1981, Buzdin1983, Machida1984, Buzdin1987}, because the order parameter $\Delta(x)$~\eqref{SCCf} depends only on $|f_{\lambda,b}^+(x)|$, and the ground-state energy can be calculated without invoking the relative phase.  However, in general, response functions involve matrix elements between various Bogoliubov amplitudes, and the relative phase is absolutely necessary for maintaining the particle-hole symmetry. Note that the solution given by Eqs.~(\ref{iden}) and (\ref{fspm}) is a particular case of a more general type of multi-periodic solutions \cite{Dubrovin1975}.

Because the pairing potential $\Delta(x)$ is periodic, the amplitudes $f_{\lambda b}^{\pm}(x)$ in Eq.~(\ref{fspm}) must satisfy the Bloch theorem $f_{\lambda,b}^\pm(x+l)=e^{ip_{\lambda,b}l}f_{\lambda,b}^\pm(x)$ with the quasi-momenta $p_{\lambda,b}$.  Here the subscript $b=\pm$ labels the two momentum branches $p_{\lambda,b}$ corresponding to a given eigenenergy $E_\lambda$.  Thus, we can write
\beg\label{Fourier}
f_{\lambda,b}^\pm(x)=e^{ip_{\lambda,b}x}\phi_{\lambda,b}^\pm(x),
\en
where $\phi_{\lambda,b}^\pm(x+l)=\phi_{\lambda,b}^\pm(x)$ is a periodic function, which can be expanded in the Fourier series
\beg\label{phims}
\phi_{\lambda,b}^\pm(x)=\sum\limits_{m=-\infty}^\infty
\tilde\phi_{\lambda,b}^\pm(m)\,e^{imQx},
\quad Q=\frac{2\pi}{l}.
\en
Using the energy-momentum relation $p_{\lambda,b}\equiv p_{b}(E_\lambda)$, we can obtain the energy dispersion $E_\lambda(p_\lambda)$ shown in Fig.~\ref{fig:exactdispersion} and the density of states per spin $\rho_b(E_\lambda)=|dp_{\lambda,b}/dE_\lambda|/2\pi$.  Below we omit the eigenstate label $\lambda$ for brevity.  First we consider the energy range $0\leq E\leq{\cal E}_2$ and find from Eq.~(\ref{fspm}) \cite{Machida1984,Mertsching1981}:
\beg\label{q2E2}
\begin{split}
p_b(E)&=-\frac{b\Delta_2}{v_F{\mathrm K}(k)}{\cal M}(\varphi_E,k'),\quad E\leq {\cal E}_2,
\end{split}
\en
where $\varphi_E=\arcsin(|E|/{\cal E}_2)$ and
\beg\label{SphiE}
{\cal M}(\varphi_E,k')=[{\mathrm E}(k)-{\mathrm K}(k)]F(\varphi_E,k')
+{\mathrm K}(k)E(\varphi_E,k')
\en
with $F(\varphi_E,k')$ and $E(\varphi_E,k')$ being the incomplete elliptic integrals of the first and the second kinds, correspondingly.  The function ${\cal M}(\varphi_E,k')$ remains positive for all values of the energy $E$.  The momenta corresponding to the band edge energy ${\cal E}_2=\Delta_2 k'$ are $p({\cal E}_2)=\pm(\pi\Delta_2/2v_F){{\mathrm K}(k)}=\pm Q/2$, as shown in Fig.~\ref{fig:exactdispersion}.

Similarly, for the energies above the upper band edge $E\geq {\cal E}_3$, we get
\beg\label{q2E3}
\begin{split}
p_b(E)=&\frac{b}{v_F}\left[\sqrt{\frac{(E^2-{\cal E}_2^2)(E^2-{\cal E}_3^2)}{E^2}}\right.\\
&\left.+\frac{\Delta_2}{{\mathrm K}(k)}{\cal M}(\tilde{\varphi}_E,k')\right], \quad E\geq {\cal E}_3,
\end{split}
\en
where $\tilde{\varphi}_E=\arcsin({\cal E}_3/|E|)$. Notice the minus sign in front of the square brackets in Eq.~(\ref{q2E3}), which indicates that the branch index $b$ actually changes across the gap, as shown in Fig.~\ref{fig:exactdispersion}.  For the energies away from the gap $E\ll{\cal E}_2$ and $E\gg{\cal E}_3$, Eqs.~(\ref{q2E2}) and (\ref{q2E3}) reproduce the original linearized dispersion relation $E=v_Fp$ in Eq.~\eqref{eq:linearized+tilde}.  [The quasi-momentum $p$, introduced in Eq.~\eqref{Fourier} and used in the rest of the paper, actually corresponds to $p-\tilde p_F$, where the latter momentum $p$ is the one used in Sec.~\ref{sec:qualitative}.]  Equations (\ref{q2E2}) and (\ref{q2E3}) determine the single-particle spectrum of the quasi-1D FFLO state, which is shown in Fig.~\ref{fig:exactdispersion}.

The single-particle density of states $\rho(E)=|dp/dE|/2\pi$ is given by the following expression \cite{Mertsching1981}
\beg\label{DOS}
\rho(E)=\rho_F\frac{|2E^2-{\cal E}_2^2-{\cal E}_3^2+\langle\Delta^2\rangle|}
{\sqrt{(E^2-{\cal E}_3^2)(E^2-{\cal E}_2^2)}}, \quad
\rho_F=\frac{1}{2\pi v_F},
\en
where $\rho_F$ is the density of states in the normal state and
\beg
\langle\Delta^2\rangle=\frac{1}{l}\int\limits_0^l\Delta^2(x)\,dx=\Delta_2^2\left(2-k^2-2\frac{\mathrm E(k)}{\mathrm K(k)}\right).
\en
Plots of the density of states for difference values of $h$ are shown in the top row in Fig.~\ref{fig:DoS}.  At the point of transition into the spatially homogeneous superfluid phase ($k\to1$ and $h\to h_c$), Eq.~(\ref{DOS}) reproduces the BCS density of states.

Having obtained the exact Bogoliubov amplitudes, we can compute the
spin density $\rho_s(x)$:
\begin{align}\label{rhos}
\rho_s(x)\equiv\langle\hat s_z(x)\rangle = \langle\psi_\uparrow^\dag(x)\psi_\uparrow(x)\rangle
- \langle\psi_\downarrow^\dag(x)\psi_\downarrow(x)\rangle,
\end{align}
where $\hat s_z(x)$ is the spin-density operator.  Using Eqs.~\eqref{Bogtrans} and \eqref{vacuum}, we find
\begin{equation}\label{up-down}
\begin{split}
\langle\psi_\sigma^\dag(x)\psi_\sigma(x)\rangle & = \sum\limits_\lambda
\left\{|U_\lambda(x)|^2 n_F(E_\lambda-\sigma h)\right.\\
&+\left.|V_\lambda(x)|^2[1-n_F(E_\lambda+\sigma h)] \right\}
\end{split}
\end{equation}
and
\begin{equation}\label{eq:rho_s}
\begin{split}
\rho_s(x)=\sum\limits_\lambda&\left[|U_\lambda(x)|^2+|V_\lambda(x)|^2\right]\times
\\&\times\left[n_F(E_\lambda-h)-n_F(E_\lambda+h)\right].
\end{split}
\end{equation}
where the sum is taken over positive $E_\lambda$, as discussed below.  Eq.~\eqref{eq:rho_s} has a  clear physical interpretation.  First, at $T=0$, the contribution to the spin density comes only from the middle band with the energies $|E_\lambda|<h$, which is occupied by the majority spin only, as shown in Fig.~\ref{fig:exactdispersion}.  Second, for a given value of $\lambda$, the spatial integral of the spin density is equal to one because of the normalization condition for the Nambu spinors~\eqref{eq:norm}.  Thus, at $T=0$, the integral
\begin{align}\label{eq:spindens_Q}
&n_s=\frac{1}{2L}\int_{-L}^L dx\,\rho_s(x)
=\frac{1}{2L}\sum_\lambda\theta(h-E_\lambda)\\
&=2\int_{-\infty}^\infty dE\,\rho(E)\,\theta(h-|E|)=4\int_0^{Q/2} \frac{dp}{2\pi}=\frac{Q}{\pi}. \nonumber
\end{align}
relates the spin imbalance $n_s$ and wavevector $Q=p^{\uparrow}_F-p^{\downarrow}_F$, as discussed in Sec.~\ref{sec:qualitative}.  The factor of 2 in the second line of Eq.~\eqref{eq:spindens_Q} comes from summation over the index $\alpha$ representing the $\pm p_F$ branches, see Fig.~\ref{fig:phspectrum}.

\begin{figure}
\includegraphics[width=0.8\linewidth]{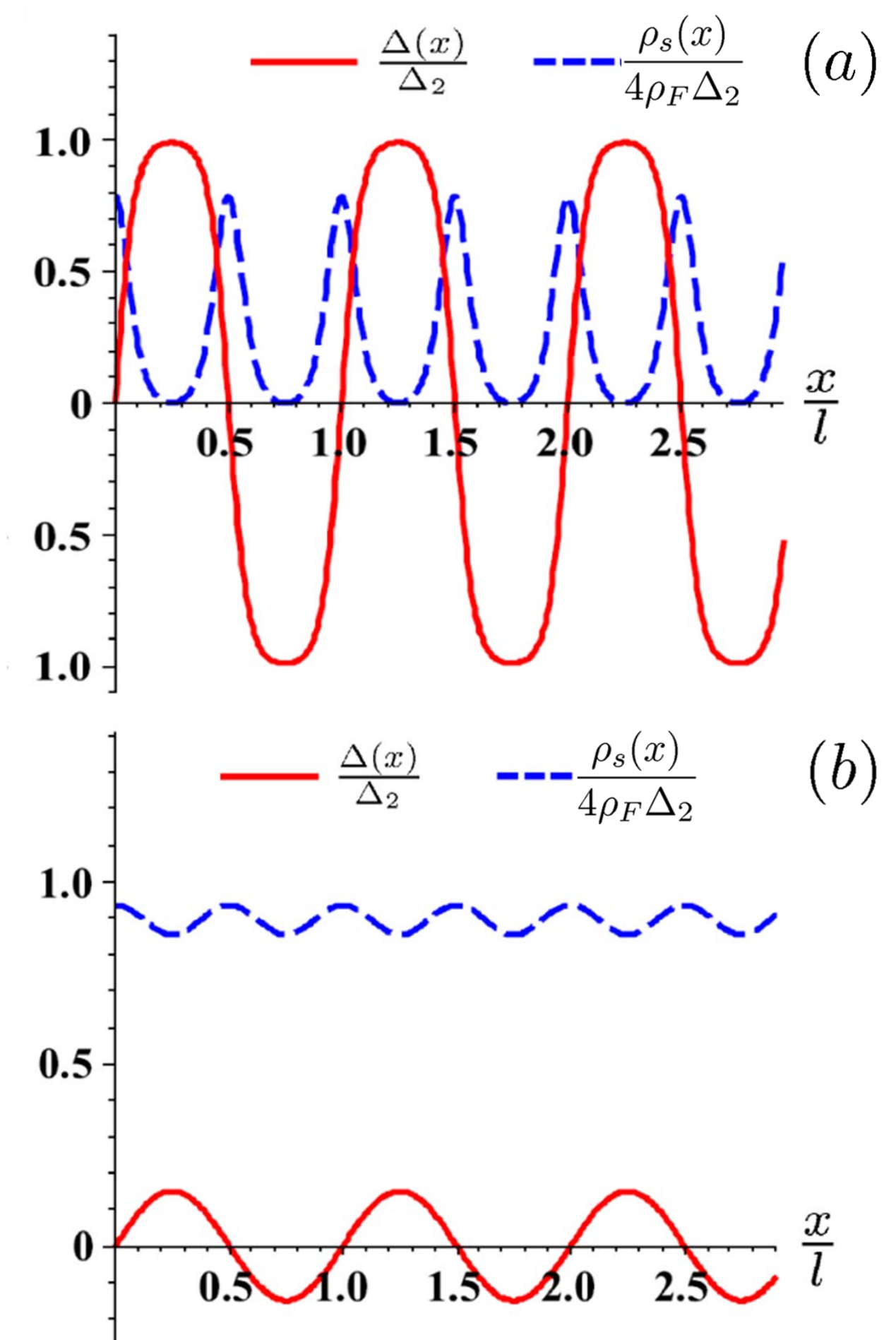}
\caption{(Color online) The pairing potential $\Delta(x)$ and the spin density $\rho_s(x)$ calculated for the quasi-1D FFLO state. (a) The soliton-lattice regime for $h$ close to the critical value $h_c$.  The spikes of $\rho_s(x)$ are aligned with the kink solitons, where $\Delta(x)$ changes sign.  (b) The Larkin-Ovchinnikov regime for $h\gg\Delta_0$.  The spin-density modulation is small compared with the uniform spin background.}
\label{fig:spin}
\end{figure}

It is instructive to consider the limit $h\rightarrow h_c$, where the imbalance corresponds to just one excessive majority atom $n_s=1/2L$.  In this regime, the order parameter forms a domain wall $\Delta(x)=\Delta_0\tanh(\Delta_0/v_Fx)$, which binds the unpaired atom.  There is a single normalizable midgap state at $E_{\lambda}=0$, which is localized at the domain wall and is characterized by the Witten topological index~\cite{Junker_book}
\begin{align}\label{eq:U0V0}
\left(\begin{array}{c}
  U_0(x) \\
  V_0(x) \\
\end{array}\right)=\sqrt{\frac{\Delta_0}{2v_F}}\frac{1}{\cosh\left(\frac{\Delta_0}{v_F}x\right)}\left(\begin{array}{c}
  \cos(\tilde p_F x) \\
  i \sin(\tilde p_F x) \\
\end{array}\right).
\end{align}
Using Eq.~\eqref{eq:U0V0}, we obtain the spin density from Eq.~\eqref{eq:rho_s}
\begin{align}
\rho_s(x)=\frac{\Delta_0}{2v_F}\frac{1}{\cosh^2(x\Delta_0/v_F)},
\end{align}
which, indeed, corresponds to the unpaired spin localized at the domain wall around $x=0$.  With the increase of the population imbalance, it becomes energetically favorable for the system to create more domain walls in order to accommodate excessive spins.  This regime corresponds to the soliton lattice.  In general, we can calculate $\rho_s(x)$ for an arbitrary population imbalance using the exact solution discussed above.  Using Eqs.~\eqref{eq:UnVn} and \eqref{fuv},  we rewrite Eq.~\eqref{eq:rho_s} at $T=0$ as
\begin{align}\label{Szgen0}
\rho_s(x)=\sum_\lambda(|f_\lambda^+(x)|^2+|f_\lambda^-(x)|^2)\,\theta(h-E_\lambda),
\end{align}
where we omitted the fast-oscillating terms with the wavevector $2p_F$.  After substituting Eqs.~\eqref{fspm} and \eqref{fspm1} into Eq.~\eqref{Szgen0} and using the identity $\sum_\lambda f(E_\lambda)=2L\int dE\,\rho(E)\,f(E)$, we find
\begin{align}\label{Szgen}
\!\!&\frac{\rho_s(x)}{4\rho_F\Delta_2}=-\int_0^{{\cal E}_2} \frac{dE}{2\Delta_2} \,
\frac{2E^2-{\cal E}^2_2-{\cal E}^2_3+\Delta^2(x)}
{\sqrt{({\cal E}^2_2-E^2)({\cal E}^2_3-E^2)}}\\
\!\!&\!=\mathrm E(k')-\frac{1\!-\!k'^2}{2}{\mathrm K(k')}
\left\{1\!+\!(1\!-\!k'^2)\text{sn}^2[(1\!+\!k')\zeta,\nu]\right\},\nonumber
\end{align}
The plots of $\Delta(x)$ and $\rho_s(x)$ are shown in Fig.~\ref{fig:spin}.  In the soliton-lattice limit depicted in Panel (a), the spin-density spikes are well pronounced, because the distance between solitons is fairly long.  With the increase of the spin imbalance, the solitons start to overlap, and the system crosses over to the LO phase with the sinusoidal $\Delta(x)$ shown in Panel (b).  The midgap states become hybridized and extended over many domain walls.  As a result, the amplitude of spin-density modulation becomes small compared with the uniform spin background, as shown in Panel (b).  Thus, it may be difficult to detect the modulation of $\rho_s(x)$ in the LO limit experimentally.

Finally, we point out that the spin density $\rho_s(x)=\rho_s(x+l/2)$ is a periodic function with the period $l/2=\pi/Q$, so it can be expanded into a Fourier series with the coefficients $\tilde\rho_s(m)$
\begin{align}
&\rho_s(x)=2\sum\limits_{m=0}^\infty \tilde\rho_s(m)\cos(2Qmx),
\label{eq:szfour}\\
&\tilde\rho_s(m)=\frac{4}{l}\int_{0}^{l/4}dx\,\rho_s(x)\,\cos(2Qmx).
\label{eq:szinvfour}
\end{align}
Because of the symmetry $\rho_s(x)=\rho_s(-x)$, the Fourier expansion has only cosine functions.  The coefficients $\tilde\rho_s(m)$ can be computed using Eq.~(\ref{Szgen})
\begin{align}\label{eq:DFT_Sz}
\frac{\tilde\rho_s(m)}{4\rho_F\Delta_2}&=\left[\mathrm E(k')-\frac{1-k'^2}{2}\mathrm K(k')\right]\delta_{m,0}\nonumber\\
&-\frac{(1-k'^2)^2}{2}\,\mathrm K(k')\,Y(k',m),
\end{align}
where the dimensionless function $Y(k',m)$ is defined as
\begin{align}
Y(k',m)=\int_0^1dz\, \text{sn}^2\left[\mathrm K(\nu)z,\nu\right]\,\cos(\pi m z)
\end{align}
with $\nu=(1-k')/(1+k')$.  The Fourier coefficients $\tilde\rho_s(m)$ can be calculated by expressing the $\rm sn$ function in terms of an infinite series and taking the spatial integral over $z$, see Ref.~\cite{Machida1984}.  We will use the expansion (\ref{eq:szfour}) in the next Section when discussing the elastic optical Bragg scattering experiments.

Here we conclude our overview of the exact mean-field pairing solution for a quasi-1D Fermi system with a spin-population imbalance.  One of the hallmarks of this solution is the midgap energy band populated by the majority spins.  In the next Section, we propose and theoretically analyze several experiments for detection of the soliton lattice.

\section{Experimental detection of the soliton lattice}\label{sec:experiment}

The soliton lattice in the quasi-1D FFLO state can be detected by various experimental techniques, such as polarization phase-contrast imaging~\cite{Hulet2009,Bradley}, Bragg diffraction~\cite{Peil, Corcovilos}, radio-frequency spectroscopy~\cite{Chin2004, Schunck2008}, quantum polarization spectroscopy~\cite{Roscilde2009} and quantum spin-noise spectroscopy~\cite{Eckert, Bruun}.  Different experimental techniques have advantages and disadvantages.  In particular, the polarization phase-contrast imaging used in the recent Rice experiment~\cite{Hulet2009} does not have sufficient spatial resolution to resolve modulation of $\rho_s(x)$.  Thus, alternative experimental probes are needed to identify the FFLO state unambiguously.  Also, it is important to map out the parameter space where the signatures of the FFLO state are most prominent.  In this Section, we propose and analyze theoretically three experimental approaches which may be the most suitable for a clear detection of the FFLO phase. Specifically, we consider the optical Bragg diffraction on spin-density modulation, the inelastic Bragg scattering, and the radio-frequency spectroscopy.

\subsection{Optical Bragg diffraction on spin-density modulation}
\label{sec:elastic}

Spatial modulations of the spin density $\rho_s(x)$ in the FFLO state can be detected using the spin-dependent Bragg scattering of light.  This method is a cold-atom analog of the elastic polarized-neutron scattering widely used in condensed-matter physics, particularly for detecting magnetism and determining the symmetry of a superconducting order parameter, see, e.g.,\ Ref.~\cite{Mazin'95}.  The optical elastic Bragg scattering was proposed in Ref.~\cite{Corcovilos} for detection of antiferromagnetism in cold atoms described by the fermionic Hubbard model, but it can be also adapted for our problem.  This method is based on the following observation.  If the frequency $\omega$ of incident light is tuned halfway between the energy distance from the level $|3\rangle$ to the two hyperfine levels $|1\rangle$ and $|2\rangle$ in Fig.~\ref{fig:dispersion}, i.e.,\ $\omega=\omega_{23}+\omega_{12}/2$, then the light couples to the population imbalance of the levels $|1\rangle$ and $|2\rangle$, i.e.,\ to the local spin density $\rho_s(x)=\rho_\uparrow(x)-\rho_\downarrow(x)$.  As explained in more detail in Appendix~\ref{app:inelastic}, this happens because the scattering matrix elements of light on the atoms in the hyperfine states $|1\rangle$ and $|2\rangle$ have opposite signs, so they produce opposite phase shifts for the light.  Because the light frequency is detuned from atomic transitions in this regime, there is no photon absorption, whereas photon scattering is maximally sensitive to the spin-density modulation.  Thus, the cross section of elastic scattering is
proportional to the static spin-structure factor $S(q_x)$~\cite{Corcovilos}, which is determined by the Fourier transform $\tilde\rho_s(q_x)$ of the spin density $\rho_s(x)$: $S(q_x)\propto|\tilde\rho_s(q_x)|^2$.  Below, we derive this relation explicitly for the three-band model shown in Fig.~\ref{fig:dispersion}.

The energy dispersion of an atom in the state $|3\rangle$ is $\epsilon_3(p)=E_3+p^2/2m$, where $E_3=\omega_{23}+\omega_{12}$ is counted from the bottom of the lowest band in Fig.~\ref{fig:dispersion}.  We assume that the state $|3\rangle$ is not populated (i.e.,\ $\mu_3=0$), and the interaction amplitudes between atoms in the state $\ket{3}$ and with atoms in the other states $\ket{1}$ and $\ket{2}$ are negligible.  Let us introduce the annihilation operators $\hat\psi_3(x)$ for an atom in the state $|3\rangle$ and $\hat a_{\bm k}$ for a photon with momentum $\bm k$.  The interaction of light with the atoms is governed by the Hamiltonian $\hat H_{\rm int}=\sum_\sigma(\hat H_\sigma+\hat H_\sigma^\dag)$, where the operator $\hat H_\sigma$ describes atomic transitions from the state $\sigma$ to the state $|3\rangle$ with absorption of a photon
\begin{align}\label{eq:H_sigma}
\hat H_\sigma = \sum_{k_x} \int_{-L}^L dx \,\Upsilon_{\bm k} \, e^{-ik_xx}
\hat a_{\bm k}  \, \hat\psi_3\dg(x) \, \hat\psi_\sigma(x).
\end{align}
The amplitudes $\Upsilon_{\bm k}$ contain microscopic information about atomic transitions, such as dipole matrix elements and light polarization.  To simplify notation, Eq.~\eqref{eq:H_sigma} is written for one tube, but summation over all tubes is implicitly assumed.

\begin{figure}
\includegraphics[height=2.8in,angle=90]{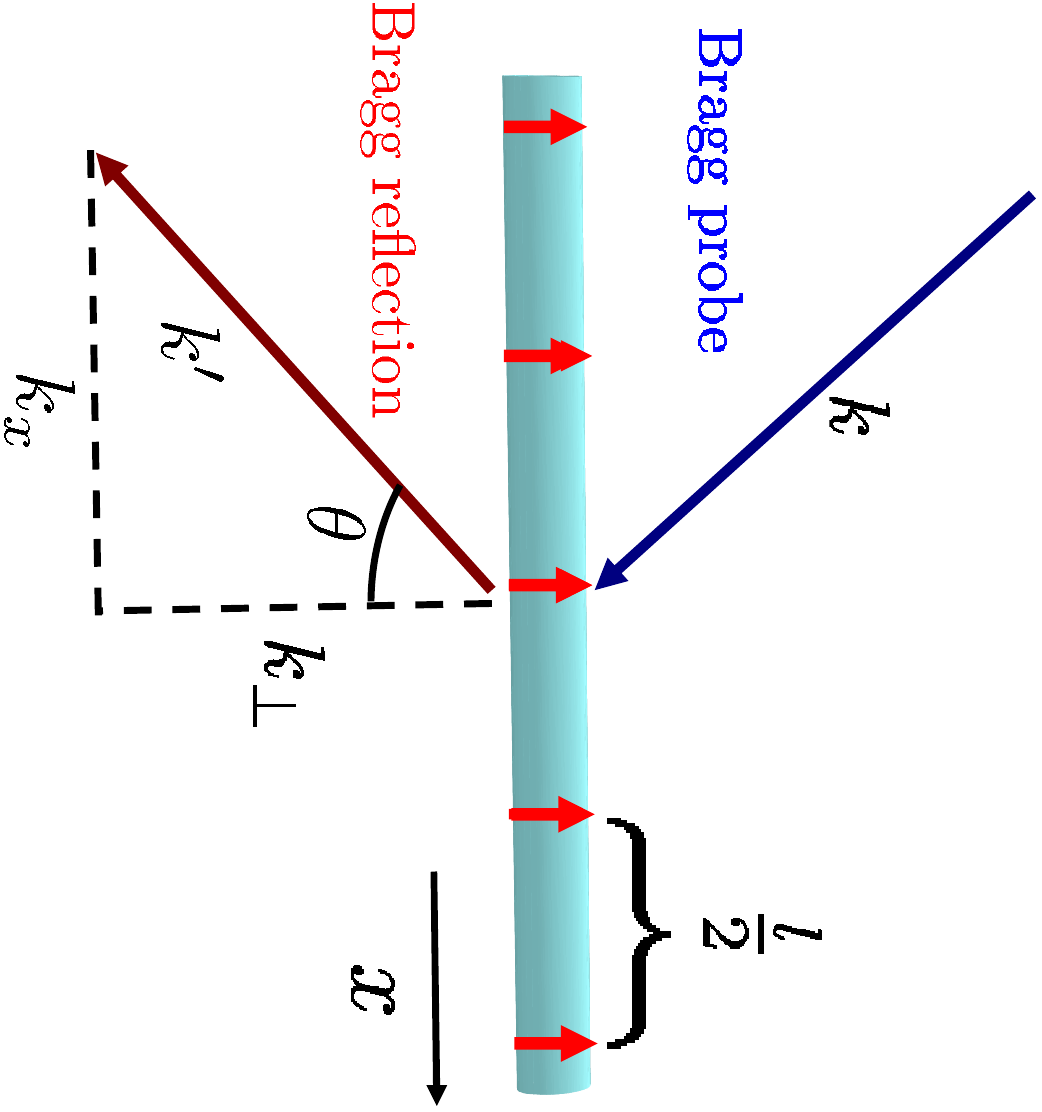}
\caption{(Color online) Schematic plot of the elastic Bragg diffraction experiment.  Photons with the momentum $|\bm k|=2\pi/\lambda$ are elastically scattered to the momentum $\bm k'$ by the spin-density modulation with the period $l/2$.  The Bragg condition for constructive interference is $l\sin\theta=m\lambda$ (or $2k_x=2Qm$), where $m$ is an integer.}
\label{fig:Bragg}
\end{figure}

We consider a process where an incoming photon with the momentum $\bm k=(k_x,\bm k_\perp)$ and energy $\omega_{\bm k}$ scatters into the state with the momentum $\bm k'=(-k_x,\bm k_\perp)$ and energy $\omega_{\bm k'}$, as shown in Fig.~\ref{fig:Bragg}.  Since $|\bm k|=|\bm k'|$, the photon energies $\omega_{\bm k}=\omega_{\bm k'}$ are the same, so the scattering process is elastic.  At the same time, the momentum transfer $q_x=k_x-k_x'=2k_x$ from the photon to the atoms in the $x$ direction allows one to probe the spatial spin-density modulation in the FFLO state.

The transition amplitude $M_{\bm k',\bm k}^{\rm el}$ for this elastic process is given by the second-order perturbation theory in $\hat H_{\rm int}$:
\begin{align}\label{eq:matrix_element}
M_{\bm k',\bm k}^{\rm el}
=\bra{0,\bm k'}\hat H_{\rm int}\frac{1}{E_{0,\bm k}-\hat H}\hat H_{\rm int}\ket{0,\bm k},
\end{align}
where $|0\rangle$ represents the ground FFLO state of the atoms in the levels $|1\rangle$ and $|2\rangle$ and the empty state of the level $|3\rangle$.  As shown in Appendix~\ref{app:inelastic}, when $\omega=\omega_{23}+\omega_{12}/2$, the transition amplitude is proportional to the Fourier transform $\tilde\rho_s(q_x)$ of the spin density:
\begin{align}\label{eq:amplitude}
M^{\rm el}_{\bm k',\bm k}=- \frac{2\,\overline\Upsilon_{\bm k}\,\Upsilon_{\bm k'}}{\omega_{12}}
\,\tilde\rho_s(k_x-k_x').
\end{align}
Then, the transition rate $W_{\bm k',\bm k}^{\rm el}$ for the elastic scattering process is given by the Fermi golden rule
\begin{align}\label{eq:probability_elastic}
& W_{\bm k',\bm k}^{\rm el}(q_x)=2\pi|M_{\bm k',\bm k}^{\rm el}|^2\,
\delta(\omega_{\bm k}-\omega_{\bm k'}) \nonumber \\
& =8\pi
\frac{|\Upsilon_{\bm k}|^2 \,|\Upsilon_{\bm k'}|^2}{|\omega_{12}|^2}
\,\tilde\rho_s^2(q_x)\,
\delta(\omega_{\bm k}-\omega_{\bm k'}),
\end{align}
where $q_x=k_x-k_x'$ is the transferred momentum.
The Fourier transform $\tilde\rho_s(q_x)$ of the spin density can be straightforwardly computed from the Fourier expansion \eqref{eq:szfour}
\begin{align}\label{eq:szkx}
&\tilde\rho_s(q_x) = \int_{-L}^{L}e^{-iq_xx} \rho_s(x)\, dx\\
&=2\pi\sum_{m=0}^{\infty}\tilde\rho_s(m)\left[\delta(q_x-2mQ)+\delta(q_x+2mQ)\right].
\label{eq:infinite}
\end{align}
We observe that the photon scattering rate $W_{\bm k',\bm k}$ peaks when the momentum transfer $q_x=k_x-k_x'$ is an even integer multiple of the momentum $Q$ of the soliton lattice.  This effect represents diffraction of light on the periodic modulation of spin density in the soliton-lattice state.  An experimental observation of this effect would give a strong evidence in favor of the spatially-inhomogeneous quasi-1D FFLO state.

\begin{figure}
\includegraphics[width=3.4in,angle=0]{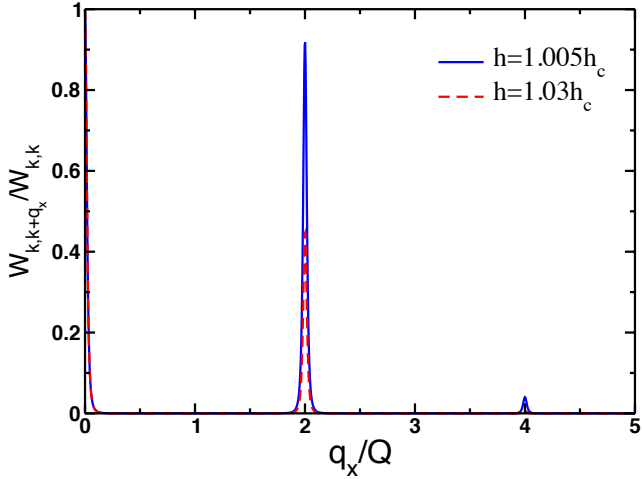}
\caption{(Color online) Transition rate $W_{\bm k',\bm k}^{\rm el}$ for the elastic Bragg scattering vs.\ the momentum transfer $q_x=k_x'-k_x$, calculated from Eq.~\eqref{eq:szkx} for $L=12l$.  The scattering rate has peaks at $q_x=2mQ$.  The effective magnetic fields $h=1.005h_c$ (solid line) and $h=1.03h_c$ (dashed line) correspond to the spin polarizations  $P\approx 4.1\%$ and $P \approx 4.2\%$ in Eq.~\eqref{eq:polarization}.  Here we used $\Delta_0/E_F=0.2$.}
\label{fig:elastic}
\end{figure}

Equation \eqref{eq:infinite} is obtained for an infinite tube with $L=\infty$, but the experimental situation~\cite{Hulet2009} corresponds to a finite system with $L/l\sim 10\div100$.  To check for finite-size effects, we calculated the integral \eqref{eq:szkx} for $L/l=12$ and plotted the results in Fig.~\ref{fig:elastic} for two values of $h$.  The peaks in the scattering rate at $q_x=2mQ$ are well-defined.  For the experimentally relevant values of $Q$,
the probe light should have the frequency $\omega$ in the visible spectrum in order to deliver the required momentum transfer $q_x=2mQ$.  This can be achieved, for example, by using the $2S_{1/2}\rightarrow{}2P_{3/2}$ transitions in $^6$Li, corresponding to the wavelength of light $\lambda=671$~nm~\cite{Corcovilos}.  Thus, although there may be additional complications due to confinement potential, our results show that the diffraction peaks are observable under realistic experimental conditions and can give a signature of the soliton-lattice formation in quasi-1D fermionic superfluids.

\subsection{Inelastic Bragg scattering}
\label{sec:inelastic}

Now we generalize the previous discussion to the inelastic Bragg scattering, where the final state of the atomic system is an excited state $\ket{\eta}$.  The transition rate is given by the Fermi golden rule:
\begin{align}\label{eq:probability_inelastic}
W_{\bm k',\bm k}^{\rm in}=2\pi\sum_{\eta}|M_{\bm k',\bm k}^{\rm in}(\eta)|^2
\,\delta(\omega_{\bm k'}-\omega_{\bm k}+E_\eta-E_0),
\end{align}
where $E_\eta-E_0$ is the excitation energy of the atomic system.  The matrix element for the inelastic process is given by the second order-perturbation theory:
\begin{align}\label{eq:amplitude_inelastic}
M_{\bm k',\bm k}^{\rm in}(\eta)=\bra{\eta,\bm k'}\hat H_{\rm int}
\frac{1}{E_{0,\bm k}-\hat H} \hat H_{\rm int}\ket{0,\bm k}.
\end{align}
Equation~\eqref{eq:matrix_element} for the elastic Bragg scattering is recovered from Eq.~\eqref{eq:amplitude_inelastic} when $\omega_{\bm k}=\omega_{\bm k'}$.  After some manipulations described in Appendix~\ref{app:inelastic}, we find
\begin{align}\label{eq:inelastic_final}
W_{\bm k',\bm k}^{\rm in}=
\frac{4|\Upsilon_{\bm k}|^2 |\Upsilon_{\bm k'}|^2}{\omega_{12}^2}
S(\omega_{\bm k}-\omega_{\bm k'},k_x-k'_x),
\end{align}
where
\begin{equation}\label{Swq}
\begin{split}
S(\Omega,q)=&\int dt\, dx_1\, dx_2 \, e^{i\Omega t -iq(x_2-x_1)}
\langle\hat{s}_z(x_2,t)\hat{s}_z(x_1,0)\rangle
\end{split}
\end{equation}
is the dynamical spin structure factor.  Here $\Omega=\omega_{\bm k}-\omega_{\bm k'}$ and $q=k_x-k'_x$ represent the energy and momentum transfers from the photon to the atoms.  When the atomic system is in the ground state, the inelastic photon scattering is only possible for $\Omega>0$, i.e.,\ when the photon loses some energy.  Equation~\eqref{eq:inelastic_final} applies when the incoming photon energy is tuned to $\omega_{\bm k}=\omega_{23}+\omega_{12}/2$, where the transition rate is maximally sensitive to the spin-spin correlations function.  For a different frequency, there are also contributions to the transition rate from the density-density correlation function~\cite{Corcovilos}.

To calculate the dynamical structure factor, we use the imaginary-time formalism and define the spin susceptibility $\chi(i\Omega_n,q)$:
\beg\label{chi}
\begin{split}
\chi(i\Omega_n,q) = & \int d\tau\, dx_1\, dx_2\, e^{i\Omega_n\tau-iq(x_2-x_1)} \\
&\times \langle \hat{T}_\tau\{\hat{s}_z(x_2,\tau)\,\hat{s}_z(x_1,0)\}\rangle,
\end{split}
\en
where $\Omega_n=2\pi nT$ is the bosonic Matsubara frequency with the temperature $T$ and an integer $n$. (We will take the limit $T\to0$ in the final results.)  In Eq. (\ref{chi}), we have introduced the notation $\hat{s}_z(x,\tau)=\exp(\hat{H}\tau)\hat{s}_z(x)\exp(-\hat{H}\tau)$.  The dynamical spin structure factor $S(\Omega,q)$ is obtained from $\chi(i\Omega_n,q)$ by the analytical continuation:
\beg\label{Swqdef}
S(\Omega,q)=-\frac{1}{\pi} {\rm Im} \, \chi(i\Omega_n\rightarrow \Omega+i\delta, q).
\en
The dynamical spin structure factor (\ref{Swq}) can be calculated either by direct evaluation of the averages using the exact expressions for the Bogoliubov amplitudes similarly to the calculation in Eq.~\eqref{up-down} or, alternatively, by employing Green's functions.  Since the latter is more general and is routinely used in computation of various response functions, we use it below.  The resulting expressions, of course, are independent of the selected method of calculation.

The spin susceptibility $\chi(i\Omega_n, q)$, Eq.~(\ref{chi}), can be expressed in terms of the normal and anomalous Green's functions for the fermions.  Details of the calculations are presented in Appendix~\ref{app:dynamic}, where we show that the spin correlation function $\chi(i\Omega_n;x,x')$ consists of two contributions from the type-I and type-II processes defined below. Performing the analytical continuation~\eqref{Swqdef}, we find the spin structure factor in the real space:
\begin{align}\label{chiS}
S(x_1,x_2;\Omega)
& = \sum\limits_{\mu\nu}
L_{\mu\nu}^{(\rm I)}(x_1,x_2) \Theta_{\mu\nu}^-
\delta(\Omega-E_\mu+E_\nu)\\
&+\sum\limits_{\mu \nu} L_{\mu\nu}^{(\rm II)}(x_1,x_2)\Theta_{\mu\nu}^+
\delta(\Omega-E_\nu-E_\mu). \nonumber
\end{align}
The sums in Eq.~\eqref{chiS} are taken over the positive energy eigenvalues  $E_{\lambda,\mu}>0$, and the occupation factors $\Theta_{\mu\nu}^{\pm}$ are
\begin{align}\label{eq:Theta-}
\Theta_{\mu\nu}^{-}&=n_F(E_\nu-h)- n_F(E_\mu-h),\\
\Theta_{\mu\nu}^{+}&=1-n_F(E_\nu+h)-n_F(E_\mu-h).
\label{eq:Theta+}
\end{align}
The delta functions in Eq.~\eqref{chiS} indicate contributions to $S(x_1,x_2;\Omega)$ from two different excitation types I and II.  The type I processes correspond to annihilation of a quasiparticle with the energy $E_\nu<{\cal E}_2$ in the midgap band and creation of a quasiparticle in the excited state with the energy $E_{\mu}>{\cal E}_3$ in the upper band in Fig.~\ref{fig:exactdispersion}.  These processes involves spin-majority quasiparticles, which occupy the midgap band, so the factor $\Theta_{\mu \nu}^-=1$ in Eq.~\eqref{eq:Theta-}, because $n_F(E_\nu-h)=1$ and $n_F(E_\mu-h)=0$.  Similar transitions for the minority spins are not possible, because they do not occupy the midgap band.  The type II processes correspond to creation of two quasiparticles with opposite spins and the energies $E_\nu$ and $E_\mu$.
The minority quasiparticle can be created either in the upper band with $E_{\nu}>{\cal E}_3$ (process IIa) or in the midgap band with $E_{\nu}< {\cal E}_2$ (process IIb), whereas the majority quasiparticle can be only created in the upper band with $E_{\mu}>{\cal E}_3$, because the midgap band is occupied by the majority spins.  In both cases, $\Theta_{\lambda\mu}^+=1$ in Eq.~\eqref{eq:Theta+}, because $n_F(E_\mu-h)=n_F(E_\nu+h)=0$.

The matrix elements $L_{\lambda\mu}^{(\rm I)}(x_1,x_2)$ and $L_{\lambda\mu}^{(\rm II)}(x_1,x_2)$ for the type I and II processes have the typical BCS structure \cite{Tinkham}:
\beg\label{SzSperp}
\begin{split}
L_{\mu \nu}^{(\rm I)}(x_1,x_2) &= \sum\limits_{\alpha,\alpha',b,b'}
P_{\mu\alpha;\nu\alpha'}^{(b,b')}(x_1)
\overline{P}_{\mu\alpha;\nu\alpha'}^{(b,b')}(x_2),
\\
L_{\mu \nu}^{(\rm II)}(x_1,x_2) &= \sum\limits_{\alpha,\alpha',b,b'}
T_{\mu\alpha;\nu\alpha'}^{(b,b')}(x_1)
\overline{T}_{\mu\alpha;\nu\alpha'}^{(b,b')}(x_2),
\end{split}
\en
where the coherence factors $P_{\lambda \alpha;\mu \alpha'}^{(b,b')}(x)$ and $T_{\lambda \alpha;\mu \alpha'}^{(b,b')}(x)$ are given by Eqs.~\eqref{eq:coherenceP} and \eqref{eq:coherenceT}.

To calculate the matrix elements, we expand the Bogoliubov amplitudes in the Fourier series as follows
\beg\label{umvm}
\left[
\begin{matrix}
U_{\lambda,1}^{(b)}(x) \\ V_{\lambda,1}^{(b)}(x)
\end{matrix}
\right]=
\sum\limits_{m=-\infty}^{\infty}
\left[
\begin{matrix}
\tilde{u}_{\lambda,b}(m) \\
\tilde{v}_{\lambda,b}(m)
\end{matrix}
\right]e^{i(p_{\lambda b}+\tilde p_F+Qm)x}.
\en
Equation~\eqref{umvm} is written for $\alpha=1$, whereas an equation for $\alpha=2$ can be obtained by interchanging $u$ and $v$ and replacing $\tilde p_F$ by $-\tilde p_F$, as follows from Eq.~\eqref{eq:UnVn}.  The amplitudes $\tilde{u}_{\lambda,b}(m)$ and $\tilde{v}_{\lambda b}(m)$ can be derived straightforwardly from Eqs.~(\ref{fuv}), (\ref{Fourier}), and (\ref{phims}).
It also follows from the particle-hole symmetry, Eq.~\eqref{reflect}, that $|\tilde u_{\lambda,-b}(m)|=|\tilde v_{\lambda b}(-m)|$.  Plots of the absolute values of several Fourier components $\tilde u_{\lambda,b}(m)$ and $\tilde v_{\lambda,b}(m)$ are shown in Appendix~\ref{app:Fourier}.

In contrast to homogeneous superfluids, the soliton lattice breaks translational symmetry.  Thus, $S(x_1,x_2; \Omega)$ and $L_{\lambda\mu}^{(\rm I,II)}(x_1,x_2)$ depend not only on the relative coordinate $x_1-x_2$, but also on the center-of-mass coordinate $(x_1+x_2)/2$.  Thus, the Fourier transforms $L_{\lambda\mu}^{(\rm I, II)}(q,K)$ depend on the two momenta $q$ and $K$ corresponding to the relative and the center-of-mass coordinates.  Substituting Eq.~\eqref{umvm} into Eqs.~\eqref{eq:coherenceP},  \eqref{eq:coherenceT}, and \eqref{SzSperp}, we obtain Eq.~(\ref{Sk}) for the Fourier transforms $L_{\lambda\mu}^{(\rm I, II)}(q,K)$.  However, to obtain the inelastic scattering rate in Eq.~\eqref{eq:inelastic_final}, we only need the dynamical spin structure factor $S(\Omega,q)$ at $K=0$ in Eq.~(\ref{Swq}).  So, we set $K=0$ in the rest of the calculations.

Because of complexity of the final equations, the frequency and momentum dependence of $S(\Omega, q)$ has to be analyzed numerically.  A technical discussion is given in Appendix~\ref{app:Fourier}.  Here we present a qualitative analysis and identify the underlying microscopic processes in the case of a small momentum transfer $|q|\ll p_F$.  It follows from Eqs. (\ref{chiS}) that
\beg\label{Supwq}
S(\Omega,q)=S_{\rm I}(\Omega,q)+
S_{\rm II}(\Omega,q),
\en
where the functions $S_{\rm I}$ and $S_{\rm II}$ are obtained from Eqs.~\eqref{Sk}, \eqref{KI}, and \eqref{KII} by setting $K=0$
\begin{align}
& S_{\rm I}(\Omega,q) =
\label{Sn} \\
& \sum\limits_{\lambda,\mu,\alpha,b,b',\{m_j\}}
{\cal K}_{\lambda\mu}^{(\rm I)}(\alpha,\alpha,b,b',\{m_j\}) \,
\delta(\Omega+E_\lambda-E_\mu)
\nonumber \\
& \times \Theta_{\lambda\mu}^{-} \delta[p_{\mu,b'}-q-p_{\lambda,b}-Q(m_1-m_1')] \, \delta_{m_1-m_1',m_2-m_2'},
\nonumber
\end{align}
\begin{align}
& S_{\rm II}(\Omega,q) =
\label{Sa} \\
& \sum\limits_{\lambda,\mu,\alpha,b,b',\{m_j\}}
{\cal K}_{\lambda\mu}^{(\rm II)}(\alpha,\bar\alpha,b,b',\{m_j\}) \,
\delta(\Omega-E_\lambda-E_\mu)
\nonumber \\
& \times \Theta_{\lambda \mu}^{+} \delta[p_{\mu,b'}-q+p_{\lambda,b}+Q(m_1+m_1')]\, \delta_{m_1+m_1',m_2+m_2'}.
\nonumber
\end{align}
Here the sums are taken over the energy level labels $\lambda$ and $\mu$, the energy branch labels $b$ and $b'$, the Fourier indices $\{m_j\}=\{m_1,m_1';m_2,m_2'\}$, and the Fermi points label $\alpha$.  The label $\alpha'=\alpha$ in Eq.~\eqref{KI} and $\alpha'\neq\alpha$ in Eq.~\eqref{KII} is selected so that $|q|\ll p_F$.

\begin{figure}
\includegraphics[width=\linewidth]{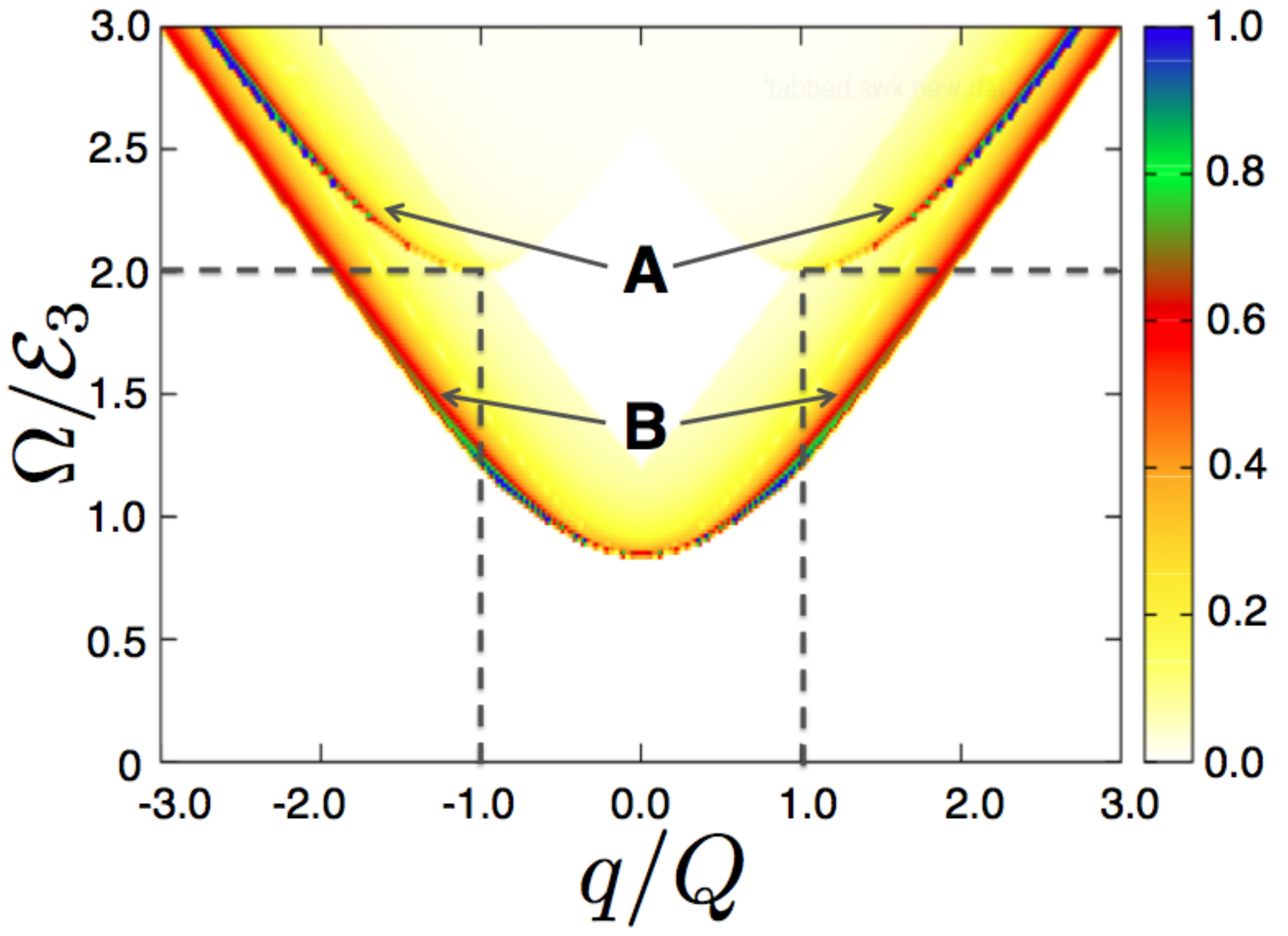}
\caption{(Color online) Contour plot of the dynamical spin structure factor $S(\Omega,q)$ vs.\ the momentum transfer $q$ and the energy-transfer $\Omega$.   Line A represents the threshold for the type IIa processes
creating two quasiparticles in the upper band in Fig.~\ref{fig:exactdispersion}.  The minimal energy transfer threshold $\Omega=2{\cal E}_3$ is achieved at $q=Q$, as indicated by the dashed lines.   Line B represents the threshold for the type I and IIb processes, which involve one quasiparticle in the midgap band and another in the upper band in Fig.~\ref{fig:exactdispersion}.  The minimal energy-transfer threshold $\Omega={\cal E}_3-{\cal E}_2$ is achieved at $q=0$.  The spin structure factor $S(\Omega,q)$ is maximal at the threshold lines A and B, as indicated by the bright colors.  The plot is calculated for the effective magnetic field $h=1.05h_c$ corresponding to the dimensionless spin polarization $P\approx 4.3\%$ at $\Delta_0/E_F=0.2$.}
\label{fig:formfactor}
\end{figure}

The numerically-calculated dynamical spin structure factor $S(\Omega,q)$ is plotted vs.\ the momentum transfer $q$ and the energy transfer $\Omega$ in Fig.~\ref{fig:formfactor} for $T=0$.  Because the experimentally measurable inelastic scattering rate in Eq.~\eqref{eq:inelastic_final} is proportional to $S(\Omega,q)$,  below we call $S(\Omega,q)$ the signal for shortness.  The signal is maximal at the lines A and B in the $(q,\Omega)$ space in Fig.~\ref{fig:formfactor}.  For a fixed $q$, the signal is strictly zero for the values of $\Omega$ below the threshold represented by line B.  For the values of $\Omega$ above line B, the signal is non-zero, but maximal signal is achieved at the threshold line itself.  Similarly, an additional sharp increase of the signal is obtained when crossing line A.

Line A in Fig.~\ref{fig:formfactor} originates from the type IIa processes, where two quasiparticles are created in the upper band in Fig.~\ref{fig:exactdispersion} with the energies $E_\mu>{\cal E}_3$ and $E_\nu>{\cal E}_3$.  In this case, the minimal energy-transfer threshold is $\Omega=2{\cal E}_3$, as shown by the horizontal dashed lines in Fig.~\ref{fig:formfactor}.  The type IIa processes are analogous to the Cooper-pair breaking in the BCS theory.  However, the unusual feature of the FFLO state is that the minimal energy-transfer threshold is achieved at the non-zero momentum transfers $q=\pm Q$, as shown by the vertical dashed lines in Fig.~\ref{fig:formfactor}.  It is a consequence of the non-zero total momenta $\pm Q$ of the Cooper pairs in the FFLO state, as discussed in Sec.~\ref{sec:qualitative}.

Line B in Fig.~\ref{fig:formfactor} originates from the type I and IIb processes, which involve destruction or creation of one quasiparticle in the midgap band with the energy $E_{\nu}<{\cal E}_2$ and creation of another quasiparticle in the upper band with the energy $E_{\mu}>{\cal E}_3$, see Fig.~\ref{fig:exactdispersion}.  Thus, the minimal energy-transfer threshold for line B is achieved at $\Omega={\cal E}_3-{\cal E}_2$, which is significantly lower than the BCS pair-breaking threshold of $2\Delta_0$ (roughly three times lower for the parameters used in Fig.~\ref{fig:formfactor}).  The presence of line B is a characteristic feature of the FFLO state.  Naively, one might expect that the excessive spin-majority fermions are unpaired, and, thus, produce a gapless spectrum of energy excitations.  However, the self-consistent solution discussed in Sec.~\ref{sec:exact} yields exactly one spin-up fermion per kink, resulting in the fully filled midgap band and the energy-transfer threshold represented by line B.

While the conservation laws of energy and momentum explain the thresholds in
Fig.~\ref{fig:formfactor}, it is not yet clear why the signal is maximal at the threshold lines.  A technical discussion of this question is given in Appendix~\ref{app:Fourier}.  Here we present a qualitative, heuristic explanation.  Although we systematically wrote all equations in the paper using positive energy eigenvalues $E_{\mu}>0$ of the BdG equation \eqref{BdG3D}, it is also possible to make a particle-hole transformation and use both positive and negative energies $E_{\mu}$.  The spin susceptibility has a simpler form given by Eq.~\eqref{eq:full_corr} in this representation, which effectively treats all excitations as the type I processes involving both positive and negative energy branches in Fig.~\ref{fig:exactdispersion}.  In order to obtain a qualitatively correct result, we restrict our consideration to the branches with positive slope in Fig.~\ref{fig:exactdispersion}, because they correctly reproduce the normal-state energy spectrum in the limit where the pairing potential vanishes.  Energy and momentum conservation in interaction between photons and Bogoliubov quasiparticles requires that the equation $E(p-q)+\Omega=E'(p)$ is satisfied.  Geometrically, the left-hand side of this equation represents the energy dispersion curve $E(p)$ of the Bogoliubov quasiparticles in an occupied band, which is shifted horizontally by the transferred momentum $q$ and vertically by the transferred energy $\Omega$.  The right-hand side of the equation represents another, unoccupied energy band $E'(p)$ of the Bogoliubov quasiparticles in Fig.~\ref{fig:exactdispersion}.  If the two curves do not cross, the equation is not satisfied, and the signal is zero.  In general, the two curves cross with non-parallel slopes at two points.  In this case, the signal is non-zero, but relatively weak, because the effective overlap volume between the two curves at the two intersection points is small.  However, when the threshold is approached, the two crossing points merge before disappearing altogether.  At the threshold, the two curves touch with parallel slopes, which greatly increases the effective overlap volume between the two curves.  Thus, we expect that the integrals of the delta functions in Eqs.~\eqref{Sn} and \eqref{Sa} would be greatly enhanced at the boundary between the domains in Fig.~\ref{fig:formfactor} where the conservation laws can and cannot be satisfied, i.e.,\ at the threshold lines A and B.  Line B represents the set of values $(q,\Omega)$ such that the displaced midgap band in Fig.~\ref{fig:exactdispersion} touches the upper band.  It is geometrically obvious that the minimal possible energy transfer is $\Omega={\cal E}_3-{\cal E}_2$ at $q=0$.  Similarly, line A represents the set of values $(q,\Omega)$ such that the displaced lower band in Fig.~\ref{fig:exactdispersion} touches the upper band.  In this case, the minimal possible energy transfer is $\Omega=2{\cal E}_3$ at $q=Q$.

These heuristic arguments indicate that the signal is enhanced by maximizing the joint density of states represented by the delta functions in Eqs.~\eqref{Sn} and \eqref{Sa}.  However, these arguments do not take into account the structure of the matrix elements in front of the delta-functions in Eqs.~\eqref{Sn} and \eqref{Sa}.  A more rigorous consideration is presented in Appendix~\ref{app:Fourier}.

We conclude this Section by emphasizing that the dynamical spin structure factor for the soliton lattice in quasi-1D fermionic superfluids has much richer structure than for conventional BCS superfluids, with several important qualitative differences discussed above.  Thus, the proposed inelastic Bragg scattering experiment to probe the dynamical spin structure factor would help to distinguish the quasi-1D FFLO state from the conventional BCS superfluid in cold-atoms experiments.

\subsection{Radio-frequency spectroscopy}
\label{sec:rf-spectroscopy}

In Secs.~\ref{sec:elastic} and \ref{sec:inelastic}, we studied the case where the frequency $\omega$ of the incoming photons is detuned significantly from the transition frequencies between the atomic energy levels in Fig.~\ref{fig:dispersion}.  In that case, the fermionic atoms are only virtually excited from the states $|1\rangle$ and $|2\rangle$ to the state $|3\rangle$, but there are no real atomic transitions, and the photons are re-emitted.  In this Section, we study the case where the incoming photons are absorbed, and the fermionic atoms make real transitions from the states $|1\rangle$ or $|2\rangle$ to the state $|3\rangle$.   Experimentally, the state $|3\rangle$ in this case is typically taken to be the hyperfine-split state of $^6$Li with $F=3/2$.
Let us define the frequency detuning $\omega_{d\sigma}$ as
\beg\label{wd}
\omega_{d\sigma}=\left\{\begin{array}{ll}
\omega-\omega_{23}-\omega_{12}, & \sigma=|1\rangle, \\
\omega-\omega_{23}, & \sigma=|2\rangle.
\end{array} \right.
\en
The interaction between the photons and the atoms is described by the operator \eqref{eq:H_sigma}.  In this Section, we study the case where the photons propagate perpendicularly to the direction of the 1D tubes, so the photon momentum $k_x=0$ is zero in the $x$ direction.  Thus, we drop the index $\bm k$ in Eq.~\eqref{eq:H_sigma} and replace $\Upsilon_{\bm k}\to\Upsilon$.  Then, Eq.~\eqref{eq:H_sigma} can be rewritten in the momentum representation, where atoms in the states $|\sigma\rangle$ and $|3\rangle$ have the same momentum $p$, i.e.,\ the atoms make vertical transitions in Fig.~\ref{fig:dispersion}:
\begin{align}\label{eq:vertical}
\hat H_\sigma = \Upsilon\sum\nolimits_p \hat a_0 \,
\hat\psi_3\dg(p) \, \hat\psi_\sigma(p).
\end{align}
We assume that the band $|3\rangle$ is empty, whereas the bands $|1\rangle$ and $|2\rangle$ are populated as shown in Fig.~\ref{fig:dispersion}.

The rate of transition is obtained using Fermi's golden rule with the perturbation $H_{\rm int}=H_{\sigma}+H_{\sigma}^\dag$ given by Eq.~\eqref{eq:vertical}.  In the case where the atoms are in the normal state, the transition rate is
\beg\label{resNorm}
W_{\sigma\to 3}^{\rm norm}(\omega_{d\sigma})
=(2L)\,(2p_F^\sigma)\,|\Upsilon|^2\,\delta(\omega_{d\sigma}).
\en
Here the factor $4Lp_F^\sigma$ represents the number of atoms in the occupied band.  The transition rate \eqref{resNorm} is proportional to $|\Upsilon|^2$, as opposed to $|\Upsilon|^4$ in Eqs.~\eqref{eq:probability_elastic} and \eqref{eq:inelastic_final} for the elastic and inelastic Bragg scattering, because the latter involves absorption and emission of a photon in the second-order perturbation theory for the Hamiltonian \eqref{eq:H_sigma}, whereas spectroscopy involves only absorption in the first-order perturbation.

In the presence of superconducting pairing, the operator $\hat\psi_\sigma(p)$ in Eq.~\eqref{eq:vertical} should be expressed in terms of the Bogoliubov operators $\hat\gamma$ and $\hat\gamma\dg$ in Eq.~\eqref{Bogtrans}.  Before studying the quasi-1D FFLO case, let us first consider a simple BCS pairing without population imbalance ($h=0$).  In this case, only the operator $\hat\gamma\dg$ has a non-zero matrix element when operating on the ground state at $T=0$, so Fermi's golden rule gives
\beg\label{WBCS}
W_{\sigma\to 3}^{\rm BCS} = 2L\,|\Upsilon|^2
\int\limits_{-\infty}^{+\infty} dp \,|\tilde V(p)|^2 \,
\delta(\xi_p+\sqrt{\xi_p^2+\Delta_0^2}-\omega_{d\sigma}).
\en
Here $\tilde V(p)$ is the Fourier transform of the amplitude $V(x)$ in Eq.~\eqref{Bogtrans}, and $\xi_p$ is the normal-state energy dispersion measured from the chemical potential in Eq.~\eqref{xi_p}.  Given that in the BCS theory
\beg\label{V-BCS}
|\tilde V(p)|^2=\frac12\left(1-\frac{\xi_p}{\sqrt{\xi_p^2+\Delta_0^2}}\right),
\en
the integral \eqref{WBCS} can be taken be changing the variable of integration to $u=\xi+\sqrt{\xi^2+\Delta_0^2}$ and introducing the density of states $\rho(\xi)=|dp/d\xi|/2\pi$.  Thus we obtain \cite{Torma2000,Dzero2007}
\beg\label{redBCS}
W_{\sigma\to 3}^{\rm BCS}(\omega_{d\sigma})=2L\pi\rho_* |\Upsilon|^2
\left(\frac{\Delta_0}{\omega_{d\sigma}}\right)^2
\theta(\omega_{d\sigma}-\omega_0),
\en
where the threshold frequency is $\omega_0=\sqrt{\tilde\mu^2+\Delta_0^2}-\tilde\mu$.  The density of states $\rho_*=\rho_F/\sqrt{1+(\omega_{d\sigma}^2-\Delta_0^2)/2\omega_{d\sigma}\tilde\mu}$ is evaluated at the value of $\xi$ that satisfies the $\delta$-function in Eq.~\eqref{WBCS}, and $\rho_F$ is the density of states at the Fermi level, see Eq.~\eqref{DOS}.

\begin{figure}
\includegraphics[width=3.4in,angle=0]{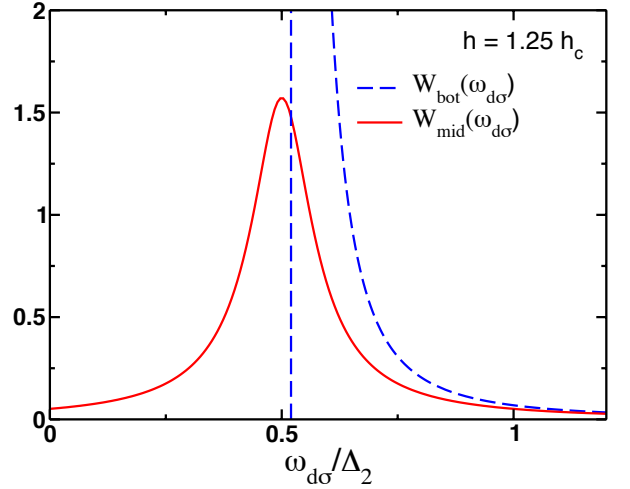}
\caption{(Color online) Absorption spectrum $W_{\sigma\to 3}^{\rm FFLO}(\omega_{d\sigma})$, Eq.~(\ref{eq:FFLO_absorption}), vs.\ the frequency detuning $\omega_{d\sigma}$, Eq.~(\ref{wd}).  For the minority fermions, $\sigma=\downarrow$, the spectrum is shown by the dashed curve and labeled as $W_{\rm bot}(\omega_{d\sigma})$, because it originates from the bottom band in Fig.~\ref{fig:phspectrum}.  It is blue-shifted similarly to the absorption spectrum (\ref{redBCS}) for the BCS state. For the majority fermions, $\sigma=\uparrow$, the spectrum consists of two terms $W_{\rm bot}(\omega_{d\sigma})$ and $W_{\rm mid}(\omega_{d\sigma})$.  The latter term is shown by the solid curve and originates from the middle band in Fig.~\ref{fig:phspectrum} with $|E_\lambda|<h$ in Fig.~\ref{fig:exactdispersion}.  The width of this peak is of the order of $\Delta_2$, and some spectral density is red-shifted to lower frequencies.  The plots are obtained for $h=1.25\,h_c$, which corresponds to the dimensionless spin polarization $P\approx 5.1\%$ assuming $\Delta_0/E_F=0.2$.}
\label{fig:rf}
\end{figure}

Physical meaning of Eq.~\eqref{redBCS} can be understood as follows.  When the superconducting gap opens, it pushes down the energy dispersion for the lower band, as illustrated in Fig.~\ref{fig:phspectrum}.  So, a higher frequency is necessary to excite an atom from the state $|\sigma\rangle$ to the state $|3\rangle$ compared with the normal-state case shown in Fig.~\ref{fig:dispersion}.  Thus, the spectral weight in Eq.~\eqref{redBCS} is blue-shifted to higher frequencies relative to Eq.~\eqref{resNorm}.  The detuning $\omega_{d\sigma}\sim\Delta_0$ corresponds to excitement of the atoms with $|p|\sim p_F$, where $\rho_*\approx\rho_F$.  On the other hand, at the threshold $\omega_{d\sigma}\to\omega_0$, the atoms are excited from the bottom of the band with $p\to0$, where the density of states diverges as $\rho_*\propto(\omega_{d\sigma}-\omega_0)^{-1/2}$.  Frequency dependence of the absorption rate \eqref{redBCS} in the BCS state is qualitatively represented by the dashed curve in Fig.~\ref{fig:rf}.  The spectral weight is blue-shifted from $\omega_{d\sigma}=0$ and monotonously decreases toward high frequencies.

Now we turn to calculation of the absorption rate for the quasi-1D FFLO state using Fermi's golden rule
\begin{align}\label{eq:FFLO_rate}
W^{\rm FFLO}_{\sigma \rightarrow 3}=2\pi \sum\nolimits_{\eta} |M(\eta, \sigma)|^2 \delta(\omega_{d\sigma} -\Delta E_{\eta, \sigma}),
\end{align}
where $M(\eta,\sigma)=\bra{\eta,0}\hat H_\sigma\ket{0,1}$ is the matrix element of a transition from the initial to the final state, and $\Delta E_{\eta,\sigma}$ is the energy difference between the final and initial atomic states. The initial state $\ket{0,1}=\ket{0}\otimes a_0^\dag \ket{0}$ corresponds to the ground state of the atoms $\ket{1}$ and $\ket{2}$ forming the FFLO superfluid and the photon state $a_0^\dag \ket{0}$.  The final state $\ket{\eta,0}$ corresponds to an excited state of the fermionic system with an atom destroyed in the state $\ket{\sigma}$ and created in the state $\ket{3}$, i.e. $\ket{\eta}=\psi_\sigma(p)\ket{0}\otimes \psi_3^\dag(p)\ket{0}$, and the photon is absorbed.   Using Eq.~\eqref{Bogtrans}, we observe that there two kinds of the final states, where a Bogoliubov quasiparticle is either destroyed $\hat\gamma_{\lambda,\sigma}\ket{0}$ or created $\hat\gamma_{\lambda,-\sigma}^\dagger\ket{0}$.  According to Eq.~\eqref{H-diagonal}, the energies of these states are $-(E_\lambda-\sigma h)$ and $E_\lambda+\sigma h$, correspondingly.  Because all energies are counted from the chemical potential $\mu_\sigma=\tilde\mu+\sigma h$, the energy of the atom in the third state effectively is $\xi_p-\sigma h$, where $\xi_p$ is defined in Eq.~\eqref{xi_p}.  Putting all the terms together, we find
\beg\label{eq:FFLO_absorption}
\begin{split}
& W_{\sigma\to 3}^{\rm FFLO}(\omega_{d\sigma}) = 2\pi |\Upsilon|^2 2L
\sum_{\alpha=1}^2 \sum_{b=\pm} \sum_{E_\lambda>0}\\
&\times \left[|\tilde U^{(b)}_{\lambda,\alpha}(p)|^2 \,
n_F(E_{\lambda}-h\sigma) \, \delta(\xi_{p} - E_\lambda - \omega_{d\sigma}) \right.\\
& + \left. |\tilde V^{(b)}_{\lambda,\alpha}(p)|^2 \, n_F(-E_{\lambda}-h\sigma)\, \delta(\xi_{p} + E_\lambda - \omega_{d\sigma}) \right] .
\end{split}
\en
where the amplitudes $\tilde V(p)$ and $\tilde U(p)$ are the Fourier transforms of the amplitudes $V_{\lambda,\alpha}^{(b)}(x)$ and $U_{\lambda,\alpha}^{(b)}(x)$ in Eq.~\eqref{umvm}, and
\begin{align} \label{p3}
& p = p_{\lambda b}+mQ-(-1)^\alpha\tilde p_F, \\
\left[ \begin{matrix}
U_{\lambda,1}^{(b)}(p) \\ V_{\lambda,1}^{(b)}(p)
\end{matrix} \right] = &
\left[ \begin{matrix}
\tilde{u}_{\lambda,b}(m) \\
\tilde{v}_{\lambda,b}(m)
\end{matrix} \right] , \;\;
\left[ \begin{matrix}
U_{\lambda,2}^{(b)}(p) \\ V_{\lambda,2}^{(b)}(p)
\end{matrix} \right]=
\left[ \begin{matrix}
\tilde{v}_{\lambda,b}(m) \\
\tilde{u}_{\lambda,b}(m)
\end{matrix} \right]. \label{Upnq}
\end{align}

Equation (\ref{eq:FFLO_absorption}) is the main result of this Subsection.  Unlike in the BCS theory, Eq.~\eqref{WBCS}, the absorption rate in the FFLO state, Eq.~\eqref{eq:FFLO_absorption}, has contributions from both $|U|^2$ and $|V|^2$.  The difference originates from the presence of the midgap band in the FFLO state in Fig.~\ref{fig:exactdispersion}.  In order to get a physical insight, let us consider the limiting case of the FFLO state with
a vanishingly small pairing potential $\Delta$ and focus on the majority fermions with $\sigma=\uparrow$.   In this case, $|U(p)|=1$ and $\xi_p=E_\lambda$ for $p>\tilde p_F$ in the first term in the sum \eqref{eq:FFLO_absorption}, so this term contributes an integral over $p$ from $\tilde p_F$ to $\tilde p_F+Q/2$ in Fig.~\ref{fig:exactdispersion}.  Similarly,
$|V(p)|=1$ and $\xi_p=-E_\lambda$ for $p<\tilde p_F$ in the second term in the sum \eqref{eq:FFLO_absorption}, so this term contributes an integral over $p$ from $\tilde p_F-Q/2$ to $\tilde p_F$ in Fig.~\ref{fig:exactdispersion}.
Together, these two terms reproduce the contribution from the midgap band to the normal-state absorption given by Eq.~\eqref{resNorm}.  On the other hand, for the minority fermions with $\sigma=\downarrow$, there is no contribution from the midgap band, because it is not populated.

Frequency dependence of the absorption rate (\ref{eq:FFLO_absorption}) is shown in Fig.~\ref{fig:rf}.  The dashed line shows $W_{\downarrow\to 3}^{\rm FFLO}(\omega_{d\sigma})$ for the minority fermions and is labeled as $W_{\rm bot}(\omega_{d\sigma})$, because the minority fermions occupy the bottom band in Fig.~\ref{fig:phspectrum}.  The dashed line is qualitatively similar to the absorption spectrum in the simple BCS case.  The spectrum is blue-shifted to higher frequencies, because the bottom band bends downward in Fig.~\ref{fig:phspectrum} due to the superconducting gap.  The contribution from the minority fermions in Fig.~\ref{fig:exactdispersion} comes from the energy branch $b=-$ with $E_\lambda>h$ and the Fourier component $\tilde v_{\lambda,-}(0)$ until the frequency $\omega_{d\sigma}$ reaches the value corresponding to the quasi-momentum at the Brillouin zone boundary $p_{\lambda,-}=-Q/2$.  At that point, the energy conservation condition enforces a  response from the quasi-particles in the energy branch $b=+$ with $E_\lambda>h$ at the zone boundary $p_{\lambda,+}=Q/2$.  To satisfy momentum conservation, the quasi-particles acquire an extra momentum $-Q$ from the soliton lattice via the Fourier component $\tilde v_{\lambda,+}(-1)$.

In contrast, the absorption rate $W_{\uparrow\to 3}^{\rm FFLO}(\omega_{d\sigma})$ for the majority fermions consists of two terms $W_{\rm bot}(\omega_{d\sigma})$ and $W_{\rm mid}(\omega_{d\sigma})$.  The first term comes from the bottom band occupied by the majority fermions in Fig.~\ref{fig:phspectrum}.  This term is the same as for the minority fermions and is shown by the dashed line in Fig.~\ref{fig:rf}.  The second term $W_{\rm mid}(\omega_{d\sigma})$ comes from the middle band in Fig.~\ref{fig:phspectrum} and Fig.~\ref{fig:exactdispersion}, which is occupied only by the majority fermions.  Because this band bends both downward and upward due to the opening of the gaps above and below it, the absorption spectrum spreads to both higher and lower frequencies.  The term $W_{\rm mid}(\omega_{d\sigma})$ is shown by the solid line in Fig.~\ref{fig:rf}.  It is important that some spectral weight for the majority fermions exhibits red shifting, in contrast to the minority fermions and the simple BCS state, which exhibit only blue shifting.  This characteristic feature can be utilized for experimental identification of the quasi-1D FFLO state.  Similar theoretical conclusions were obtained in Ref.~\cite{Torma2008}.

Absorption spectrum for the fermionic atoms was studied experimentally in Refs.~\cite{Chin2004,Schunck2008}.  The spectrum in the normal state consists of a single broadened peak, which is centered at the atomic transition frequency, i.e.,\ at $\omega_{d\sigma}=0$.  When the Fermi gas is cooled to degeneracy, an extra peak emerges at higher frequencies, followed by disappearance of the normal peak at the lowest temperatures \cite{Chin2004,Schunck2008}.  Absorption spectrum has a characteristic threshold at the frequency $\simeq\Delta_0^2/2E_F$.  These features were interpreted as the onset of the pairing gap in the spectrum of single-particle excitations \cite{Chin2004,Torma2000,Kinnunen2004,QChen2005}.  However, we are not aware of
experiments studying optical absorption spectra for the quasi-1D Fermi gases with population imbalance in the pairing regime corresponding to the FFLO state, which would be very interesting.

\section{Conclusions}  \label{sec:discussion}

In this paper, we study the quasi-1D superfluid fermionic condensate with population imbalance.  We analyze physical properties of this phase using the exact mean-field solution corresponding to the soliton lattice.  This mean-field approach is valid when the inter-tube tunneling amplitude $t_{\perp}$ is sufficiently large, so that $E_F \gg t_{\perp} \gg T_c \gg t_{\perp}^2/E_F$.  We believe these conditions can be satisfied experimentally by choosing an appropriate optical lattice depth in the transverse direction, as well as the strength of the fermion interaction.  Using the exact results for the soliton lattice, we propose and analyze several experiments aimed at detection of the exotic FFLO state in the quasi-1D cold-atom settings. First, we propose to use the optical elastic Bragg scattering to measure the spin-density modulation  accompanying formation of the soliton lattice.  Second,  the optical inelastic Bragg scattering can probe the frequency and momentum dependence of the spin-spin correlation function and provide information about the quasiparticle spectrum.  For both experiments, we identify qualitative signatures of inhomogeneous superfluidity in the quasi-1D FFLO state.  Third, we study the difference (red vs.\ blue shift) in the absorption spectra for the radio-frequency spectroscopy of the majority and minority atoms.  This difference is a characteristic feature of the FFLO state in contrast to the conventional BCS state.  Our predictions of various physical properties of the FFLO state should help to identify this exotic phase in the ongoing experiments in cold atomic gases.

In this paper, we treated the tubes in Fig.~\ref{fig:3Dtubes}a as being uniform in the $x$ direction.  However, a more realistic consideration should include the effect of a confining potential along the tube direction, which makes the problem much more complicated.  Phase diagram in the quasi-1D geometry and phase separation due to the confining potential were studied numerically in Ref.~\cite{Parish2007} using a mean-field theory.  The results of Ref.~\cite{Parish2007} for a fixed small $t_{\perp}$ indicate that the FFLO superfluid is located at the center of the trap, similarly to the strictly 1D case~\cite{Hulet2009}. Therefore, we argue that the quasi-1D regime is the most promising for observation of the FFLO physics, because the atomic motion is largely one-dimensional while maintaining advantage of the quasi-1D Fermi surface nesting.  At the same time, phase fluctuations of the pairing potential are suppressed due to the presence of weak tunneling between the tubes, which stabilizes the true long-range order.  Phase diagram as a function of the interchain hopping, as well as the dependence of the spin polarization on $t_{\perp}$ are open questions at the moment and require a more systematic study.

\section{Acknowledgements}

We thank R.~Hulet, W.~Vincent Liu and F.~Heidrich-Meisner for stimulating discussions.  This work was supported by the NSF-PFC grant to JQI.  M.D.\ acknowledges partial financial support from the Ohio Board of Regents Grant OBR-RIP-220573. R.L.\ and M.D.\ acknowledge hospitality of the Aspen Center for Physics supported by NSF grant \#1066293, where part of this work was done.

\begin{appendix}
\begin{widetext}

\section{Bragg scattering rate of light and spin structure factor for the atoms}
\label{app:inelastic}

In this Appendix, we show how the transition rate for the Bragg scattering of light is related to the spin structure factor for the atoms, both for elastic and inelastic scattering.  Using Eqs.~\eqref{eq:H_sigma} and \eqref{eq:amplitude_inelastic}, we obtain the following expression for the matrix element of transitions from the state $\ket{0,\bm k}$ representing the ground state of the atoms and a photon with the momentum $\bm k$ to the state $\ket{\eta,\bm k'}$ representing an excited state of the atoms and a photon with the momentum $\bm k'$:
\begin{align}\label{eq:M}
M_{\bm k',\bm k}^{\rm in}(\eta) = \sum_\sigma \bra{\eta,\bm k'} \hat H_\sigma^\dag
\frac{1}{E_{0,\bm k}-\hat H} \hat H_\sigma \ket{0,\bm k}.
\end{align}
The atomic level $|3\rangle$ is empty for both initial and final states.  Using Eqs.~\eqref{eq:H_sigma}, we rewrite Eq.~\eqref{eq:M} in terms of the intermediate states $|p\rangle$, where the photon is absorbed, and an atom is excited from the level $\sigma$ to the level $|3\rangle$ with the momentum $p$
\begin{align}\label{eq:Mp}
M_{\bm k',\bm k}^{\rm in}(\eta) = \sum_{\sigma,p}\int dx\, dx'\,
\overline\Upsilon_{\bm k'} \Upsilon_{\bm k}\,
e^{-ik_x'x'+ik_xx} \,
\bra{\eta} \hat\psi_\sigma\dg(x') \hat\psi_3(x') |p\rangle \,
\frac{1}{E_{0,\bm k}-E_{\sigma,p}} \,
\langle p| \hat\psi_3\dg(x) \hat\psi_\sigma(x) \ket{0}.
\end{align}
To estimate the energy difference in the denominator of Eq.~\eqref{eq:Mp}, we use the energy $E_3+p^2/2m$ for the atom promoted to the intermediate state from the initial state with the energy $E_\sigma+\varepsilon(p-k_x)$, where we took into account momentum conservation.  However, given that the energy corrections to the quasi-particle dispersion in the FFLO state and the momentum transfer from the photon are small compared with the atomic energy difference $\omega_{12}$, we can approximately use the normal-state dispersion law $E_\sigma+p^2/2m$ for the annihilated atomic state.  Then, the energy denominator becomes $\omega_{\bm k}+E_\sigma-E_3$, where $p^2/2m$ has canceled out, and $\omega_{\bm k}$ is the photon frequency.  When the photon frequency is tuned halfway between the energy distance from $|3\rangle$ to $|2\rangle$ and $|1\rangle$ (see Fig.~\ref{fig:dispersion}), i.e.,\ $\omega_{\bm k}=\omega_{23}+\omega_{12}/2$, then the denominator in Eq.~\eqref{eq:Mp} is equal to $\mp\omega_{12}/2$ for $\sigma=\uparrow,\downarrow$.  Using the plane-wave expansion $\hat\psi_3(x)=\sum_p e^{ipx}\,\hat\psi(p)/\sqrt{2L}$ and taking into account that $\langle p|\hat\psi^\dag(p)|0\rangle=\langle 0|\hat\psi(p)|p\rangle=1$, we find from Eq.~\eqref{eq:Mp}
\begin{align}\label{eq:Mxx'}
M_{\bm k',\bm k}^{\rm in}(\eta) = -\frac{2\overline\Upsilon_{\bm k'}
\Upsilon_{\bm k}}{2L\,\omega_{12}}
\sum_{p}\int dx\, dx'\,\, e^{-ik_x'x'+ik_xx} \, e^{ip(x-x')}
\bra{\eta}\hat\psi_\uparrow\dg(x')\hat\psi_\uparrow(x)
-\hat\psi_\downarrow\dg(x')\hat\psi_\downarrow(x) \ket{0}.
\end{align}
The sum over $p$ in Eq.~\eqref{eq:Mxx'} gives the delta function $\delta(x-x')$, which is then eliminated by integration over $x'$.  Thus we find
\begin{align}\label{eq:Ms}
M_{\bm k',\bm k}^{\rm in}(\eta) = -\frac{2\overline\Upsilon_{\bm k'}
\Upsilon_{\bm k}}{\omega_{12}}
\int dx\, e^{i(k_x-k_x')x}\bra{\eta}\hat s_z(x)\ket{0},
\end{align}
where $\hat s_z(x)=\hat\psi_\uparrow\dg(x)\hat\psi_\uparrow(x)-\hat\psi_\downarrow\dg(x)\hat\psi_\downarrow(x)$ is the spin-density operator.  If $|\eta\rangle$ is taken to be the ground state $|0\rangle$ in Eq.~\eqref{eq:Ms}, then $\langle0|\hat s_z(x)|0\rangle=\rho_s(x)$ as in Eq.~\eqref{rhos}, and Eq.~\eqref{eq:Ms} reproduces Eq.~\eqref{eq:amplitude} for the elastic Bragg scattering.

Substituting Eq.~\eqref{eq:Ms} into Eq.~\eqref{eq:probability_inelastic} and denoting $q=k_x-k_x'$,
we find the transition rate for the inelastic Bragg scattering
\begin{align}\label{eq:Win}
W_{\bm k',\bm k}^{\rm in}=2\pi
\frac{4|\Upsilon_{\bm k}|^2|\Upsilon_{\bm k'}|^2}{\omega_{12}^2}
\sum_\eta \int dx_1 \, dx_2 \, \bra{0}\hat s_z(x_2)\ket{\eta} \,
\bra{\eta}\hat s_z(x_1)\ket{0} \, e^{iq(x_1-x_2)}
\, \delta(\Omega -\Delta E_\eta),
\end{align}
where $\Omega=\omega_{\bm k}-\omega_{\bm k'}$ and $\Delta E_\eta=E_\eta-E_0$.  Using the identity for the delta function $\delta(\Omega-\Delta E_\eta)=\int
e^{i(\Omega-\Delta E_\eta)t}\,dt/2\pi$, the transition rate \eqref{eq:Win} can be written as
\begin{align}\label{eq:W-S}
&W_{\bm k',\bm k'}^{\rm in}=2\pi
\frac{4|\Upsilon_k|^2|\Upsilon_{k'}|^2}{\omega_{12}^2}
\int_{-\infty}^{\infty}\frac{dt}{2\pi}\int dx_1 d x_2 e^{i\Omega t}\,e^{iq(x_1-x_2)}
\sum_\eta \,
\bra{0}e^{iE_0t}\hat s_z(x_2)e^{-iE_\eta t}\ket{\eta} \,
\bra{\eta}\hat s_z(x_1)\ket{0}\\
&=\frac{4|\Upsilon_k|^2|\Upsilon_{k'}|^2}{\omega_{12}^2}
\int dt\, dx_1\, dx_2\, e^{i\Omega t+iq(x_1-x_2)}\,
\bra{0}\hat s_z(x_2,t)\hat s_z(x_1,0)\ket{0}
=\frac{4|\Upsilon_k|^2|\Upsilon_{k'}|^2}{\omega_{12}^2}
S(\omega_{\bm k}-\omega_{\bm k'},k_x-k'_x). \nonumber
\end{align}
Equation \eqref{eq:W-S} reproduces Eqs.~\eqref{eq:inelastic_final} and \eqref{Swq}.

\section{Derivation of the spin-spin correlation function}
\label{app:dynamic}

In this Appendix, we present a detailed derivation of the expression the spin-spin correlation function in Eq.~(\ref{chi}):
\beg\label{chixt}
\chi(x,\tau)=\langle\hat{T}_\tau\{\hat{s}_z(x,\tau)\hat{s}_z(0,0)\}\rangle,
\quad \hat{s}_z(x,\tau)=\hat{\psi}_\up\dg(x,\tau)
\hat{\psi}_\up(x,\tau)-\hat{\psi}\dg_\dn(x,\tau)\hat{\psi}_\dn(x,\tau),
\en
where $\hat{\psi}_\sigma(x,\tau)$ are the fermionic field operators in the Heisenberg representation. By employing standard methods of the many-body theory \cite{AGD}, one can express the function $\chi(x,\tau)$ in terms of the normal and anomalous Green's functions for the fermions, which we define as follows:
\begin{eqnarray}
&{\cal G}_\up(x,\tau;x',\tau')=-\langle\hat{T}_\tau\{\hat{\psi}_{\up}(x,\tau)\hat{\psi}_{\up}\dg(x',\tau')\}\rangle, \quad {\cal G}_\dn(x,\tau;x',\tau')=-\langle\hat{T}_\tau\{\hat{\psi}_{\dn}(x,\tau)\hat{\psi}_{\dn}\dg(x',\tau')\}\rangle, \label{Gnorm} \\ &{\cal F}_{\up\dn}(x,\tau;x',\tau')=\langle\hat{T}_\tau\{\hat{\psi}_{\up}(x,\tau)\hat{\psi}_{\dn}(x',\tau')\}\rangle,
\quad {\cal F}_{\dn\up}\dg(x,\tau;x'\tau')=\langle\hat{T}_\tau\{\hat{\psi}_{\dn}\dg(x,\tau)\hat{\psi}_{\up}\dg(x',\tau')\}\rangle.
\label{Ganom}
\end{eqnarray}
Using Wick's theorem \cite{AGD} and going into the Matsubara frequency representation, we find that the spin-spin correlation function (\ref{chixt}) can be written as a sum of two terms
\beg\label{chiST}
\chi(x_1,x_2;i\Omega_n)=\chi^{(\uparrow)}(x_1,x_2;i\Omega_n)+\chi^{(\downarrow)}(x_1,x_2;i\Omega_n)
\en
corresponding to the majority and minority fermions.  The functions $\chi^{(\up,\dn)}(x_1,x_2;i\Omega_n)$ in Eq.~(\ref{chiST}) are given by
\beg\label{chiSpm}
\begin{split}
\chi^{(\uparrow)}(x_1,x_2;i\Omega_n)&=T\sum\limits_{i\varpi}
\left[{\cal G}_{\up}(x_1,x_2;i\varpi+i\Omega_n){\cal G}_{\up}(x_2,x_1;i\varpi)+{\cal F}_{\dn\up}\dg(x_1,x_2;i\varpi+i\Omega_n)
{\cal F}_{\up\dn}(x_2,x_1;i\varpi)
\right], \\
\chi^{(\downarrow)}(x_1,x_2;i\Omega_n)&=T\sum\limits_{i\varpi}
\left[
{\cal G}_{\dn}(x_1,x_2;i\varpi+i\Omega_n){\cal G}_{\dn}(x_2,x_1;i\varpi)+{\cal F}_{\up\dn}\dg(x_1,x_2;i\varpi+i\Omega_n)
{\cal F}_{\dn\up}(x_2,x_1;i\varpi)
\right],
\end{split}
\en
where the sums are taken over the fermionic Matsubara frequency $\varpi$.
To evaluate the Matsubara sums, it is convenient to express Green's functions in terms of the Bogoliubov amplitudes given by Eqs.~(\ref{Bogtrans}).  After some algebra, we obtain the following expressions for the normal Green's functions:
\beg\label{GFn}
\begin{split}
&{\cal G}_{\up}(x_1,x_2;i\varpi)=\sum\limits_{\mu,\alpha,b}\left(\frac{U_{\mu \alpha}^{(b)}(x_1) \overline{U}_{\mu \alpha}^{(b)}(x_2)}{i\varpi-E_\mu+h}+\frac{\overline{V}_{\mu \alpha}^{(b)}(x_1)V_{\mu \alpha}^{(b)}(x_2)}{i\varpi+E_\mu+h}\right), \\ &{\cal G}_{\dn}(x_1,x_2;i\varpi)=\sum\limits_{\mu,\alpha,b}\left(\frac{U_{\mu \alpha}^{(b)}(x_1)\overline{U}_{\mu \alpha}^{(b)}(x_2)}{i\varpi-E_\mu-h}+\frac{\overline{V}_{\mu \alpha}^{(b)}(x_1)V_{\mu \alpha}^{(b)}(x_2)}{i\varpi+E_\mu-h}\right).
\end{split}
\en
Here the summation is performed over all eigenenergies $E_\mu$, the energy branch label $b$, and the right and left Fermi points label $\alpha$.  Similarly, for the anomalous Green's functions, we obtain
\beg\label{GFa}
\begin{split}
&{\cal F}_{\up\dn}(x_1,x_2;i\varpi)=\sum\limits_{\mu,\alpha,b}\left(\frac{U_{\mu \alpha}^{(b)}(x_1)\overline{V}_{\mu \alpha}^{(b)}(x_2)}{i\varpi-E_\mu+h}-
\frac{\overline{V}_{\mu \alpha}^{(b)}(x_1)U_{\mu \alpha}^{(b)}(x_2)}{i\varpi+E_\mu+h}\right), \\
&{\cal F}_{\dn\up}(x_1,x_2;i\varpi)=\sum\limits_{\mu,\alpha,b}\left(\frac{U_{\mu \alpha}^{(b)}(x_1)\overline{V}_{\mu \alpha}^{(b)}(x_2)}{i\varpi-E_\mu-h}-
\frac{\overline{V}_{\mu \alpha}^{(b)}(x_1)U_{\mu \alpha}^{(b)}(x_2)}{i\varpi+E_\mu-h}\right), \\
&{\cal F}_{\up\dn}\dg(x_1,x_2;i\varpi)=\sum\limits_{\mu,\alpha,b}\left(\frac{V_{\mu \alpha}^{(b)}(x_1)\overline{U}_{\mu \alpha}^{(b)}(x_2)}{i\varpi-E_\mu-h}-\frac{\overline{U}_{\mu \alpha}^{(b)}(x_1)V_{\mu \alpha}^{(b)}(x_2)}{i\varpi+E_\mu-h}\right),\\
&{\cal F}_{\dn\up}\dg(x_1,x_2;i\varpi)=\sum\limits_{\mu,\alpha,b}\left(\frac{V_{\mu \alpha}^{(b)}(x_1)\overline{U}_{\mu \alpha}^{(b)}(x_2)}{i\varpi-E_\mu+h}-\frac{\overline{U}_{\mu \alpha}^{(b)}(x_1)V_{\mu \alpha}^{(b)}(x_2)}{i\varpi+E_\mu+h}\right).
\end{split}
\en
In contrast to the conventional BCS theory, the anomalous Green's functions
${\cal F}_{\up\dn}(x_1,x_2;i\varpi)$ and ${\cal F}_{\dn \up}(x_1,x_2;i\varpi)$ are not equal and can be related by replacing $h\to-h$.  Substituting Eq.~\eqref{GFa} into Eq.~\eqref{eq:SCF}, we recover the self-consistency condition Eq.~(\ref{SCC})
\beg\label{selfeq2}
\overline\Delta(x)=-gT\sum_{i\varpi}{\cal F}_{\dn\up}\dg(x,x;i\varpi)e^{i\varpi 0_+}
=-2g\sum_\lambda v_\lambda(x)\,\overline{u}_\lambda(x)
\left[n_F(E_\lambda+h)-n_F(h-E_\lambda)\right].
\en
We could have employed the particle-hole symmetry relations (\ref{reflect}) and reduced the sums to the positive eigenenergies only.  However, since the correlation functions (\ref{GFn}) and (\ref{GFa}) acquire the extra prefactor of two under the transformation (\ref{reflect}), we will do it at the end of the calculation.

We now proceed with the calculation of the correlation function $\chi^{(\up)}(x_1,x_2;i\Omega_n)$. Substituting Eqs.~(\ref{GFn}) and (\ref{GFa}) into Eq.~(\ref{chiSpm}) and using the Poisson summation formula
\beg\label{Poiss}
T\sum\limits_{n=-\infty}^\infty\frac{1}{[i\pi T(2n+1)-a][i\pi T(2n+1)-b]}
=\frac{n_F(a)-n_F(b)}{a-b},
\en
where $n_F(a)$ is the Fermi distribution function, we obtain the following expression for $\chi^{(\up)}(x_1,x_2;i\Omega_n)$:
\begin{align}
\chi^{(\up)}(x_1,x_2;i\Omega_n)&=\sum\limits_{\mu\lambda;bb';\alpha\alpha'} \overline{U}_{\mu \alpha}^{(b)}(x_2) U_{\nu \alpha'}^{(b')}(x_2) \left[{U}_{\mu \alpha}^{(b)}(x_1)\overline{U}_{\nu \alpha'}^{(b')}(x_1)+ {V}_{\mu \alpha}^{(b)}(x_1)\overline{V}_{\nu \alpha'}^{(b')}(x_1)\right]\frac{n_F(E_\mu-h)-n_F(E_\nu-h)}{-i\Omega_n+E_\mu-E_\nu}\nonumber\\
&+\sum\limits_{\mu\lambda;bb';\alpha\alpha'} V_{\mu \alpha}^{(b)}(x_2) \overline{V}_{\nu \alpha'}^{(b')}(x_2) \left[\overline{V}_{\mu \alpha}^{(b)}(x_1)V_{\nu \alpha'}^{(b')}(x_1)+\overline{U}_{\mu \alpha}^{(b)}(x_1)U_{\nu \alpha'}^{(b')}(x_1)\right]\frac{n_F(-E_\mu-h)-n_F(-E_\nu-h)}{-i\Omega_n-E_\mu+E_\nu}\nonumber\\
&+\sum\limits_{\mu\lambda;bb';\alpha\alpha'} \overline{U}_{\mu \alpha}^{(b)}(x_2) \overline{V}_{\nu \alpha'}^{(b')}(x_2) \left[{U}_{\mu \alpha}^{(b)}(x_1)V_{\nu \alpha'}^{(b')}(x_1)-{V}_{\mu \alpha}^{(b)}(x_1)U_{\nu \alpha'}^{(b')}(x_1)\right]\frac{n_F(E_\mu-h)-n_F(-E_\nu-h)}{-i\Omega_n+E_\mu+E_\nu}\nonumber\\
&+\sum\limits_{\mu\lambda;bb';\alpha\alpha'} {V}_{\mu \alpha}^{(b)}(x_2){U}_{\nu \alpha'}^{(b')}(x_2) \left[\overline{V}_{\mu \alpha}^{(b)}(x_1)\overline{U}_{\nu \alpha'}^{(b')}(x_1)-\overline{U}_{\mu \alpha}^{(b)}(x_1)\overline{V}_{\nu \alpha'}^{(b')}(x_1)\right]\frac{n_F(-E_\mu-h)-n_F(E_\nu-h)}{-i\Omega_n-E_\mu-E_\nu}
\end{align}
The correlation function $\chi^{(\dn)}(x_1,x_2;i\Omega_n)$ can be obtained by replacing $h\rightarrow -h$. To simplify the presentation, it is convenient to introduce the following notation for the matrix elements:
\begin{align}\label{eq:coherenceP}
&P_{\mu \alpha;\nu \alpha'}^{(b,b')}(x)=
{U}_{\mu,\alpha}^{(b)}(x)\overline{U}_{\nu,\alpha'}^{(b')}(x)+
{V}_{\mu,\alpha}^{(b)}(x)\overline{V}_{\nu,\alpha'}^{(b')}(x),\\
&T_{\mu \alpha;\nu \alpha'}^{(b,b')}(x)=
V_{\mu,\alpha}^{(b)}(x)U_{\nu,\alpha'}^{(b')}(x)-U_{\mu,\alpha}^{(b)}(x)V_{\nu,\alpha'}^{(b')}(x).
\label{eq:coherenceT}
\end{align}
Under the particle-hole symmetry $(E_\mu,E_\nu)\rightarrow (-E_{\mu},-E_\nu)$, one can show that $P_{\mu \alpha;\nu \alpha'}^{(b,b')}(x)\rightarrow \overline{P}_{\mu \alpha;\nu \alpha'}^{(b,b')}(x)$ and $T_{\mu \alpha;\nu \alpha'}^{(b,b')}(x)\rightarrow \overline{T}_{\mu \alpha;\nu \alpha'}^{(b,b')}(x)$. Also,  for $(E_\mu, E_\nu) \rightarrow (-E_{\mu},E_\nu)$ and $(E_\mu, E_\nu) \rightarrow (E_{\mu}, - E_\nu)$, the matrix elements transform as $P_{\mu \alpha;\nu \alpha'}^{(b,b')}(x)\rightarrow - \overline{T}_{\mu \alpha;\nu \alpha'}^{(b,b')}(x)$ and $P_{\mu \alpha;\nu \alpha'}^{(b,b')}(x)\rightarrow T_{\mu \alpha;\nu \alpha'}^{(b,b')}(x)$, respectively.

Using the symmetry properties of the matrix elements, we can simplify the expressions for the correlation function.  Exchanging the indices $\{\mu,\alpha,b\}\leftrightarrow \{\nu,\alpha',b'\}$ in the expression for $\chi^{(\dn)}(x_1,x_2;i\Omega_n)$ and adding it to $\chi^{(\up)}(x_1,x_2;i\Omega_n)$, we finds
\begin{align}\label{eq:full_corr}
\chi(x_1,x_2;i\Omega_n)&=4\sum\limits_{\mu\lambda;bb';\alpha\alpha'} P_{\mu \alpha;\nu \alpha'}^{(b,b')}(x_1)\overline{P}_{\mu \alpha;\nu \alpha'}^{(b,b')}(x_2)\frac{n_F(E_\mu-h)-n_F(E_\nu-h)}{-i\Omega_n+E_\mu-E_\nu}.
\end{align}
Here the factor of $4$ comes from combining the terms related by the particle-hole symmetry. Also, note that the sums in Eq.~\eqref{eq:full_corr} are taken over both positive and negative energies $E_\lambda$.  Performing the analytical continuation $i\Omega_n \rightarrow \Omega+i \delta$ and taking the imaginary part of the spin-spin susceptibility as in Eq.~\eqref{Swqdef}, we obtain the spin structure factor in the real space
\begin{align}
S(x_1,x_2,\Omega)=4\sum\limits_{\mu\lambda;bb';\alpha\alpha'} P_{\mu \alpha;\nu \alpha'}^{(b,b')}(x_1)\overline{P}_{\mu \alpha;\nu \alpha'}^{(b,b')}(x_2)\left[n_F(E_\nu-h)-n_F(E_\mu-h)\right]\delta(\Omega-E_\mu+E_\nu).
\end{align}
In order to understand various processes contributing the spin structure factor, it is instructive to consider the cases $E_\mu >0$, $E_\nu >0$ and $E_\mu >0$, $E_\nu <0$ separately (the other cases do not contribute to $S(x_1,x_2,\Omega)$ for $h>0$ and $\Omega >0$):
\begin{align}\label{chiSn}
S^{(\rm I)}(x_1,x_2,\Omega)&=4\sum\limits_{\mu>0, \nu>0;bb';\alpha\alpha'} P_{\mu \alpha;\nu \alpha'}^{(b,b')}(x_1)\overline{P}_{\mu \alpha;\nu \alpha'}^{(b,b')}(x_2)\left[n_F(E_\nu-h)-n_F(E_\mu-h)\right]\delta(\Omega-E_\mu+E_\nu),\\
S^{(\rm II)}(x_1,x_2,\Omega)&=4\sum\limits_{\mu>0, \nu>0;bb';\alpha\alpha'} T_{\mu \alpha;\nu \alpha'}^{(b,b')}(x_1)\overline{T}_{\mu \alpha;\nu \alpha'}^{(b,b')}(x_2)\left[1-n_F(E_\nu+h)-n_F(E_\mu-h)\right]\delta(\Omega-E_\mu-E_\nu).\label{chiSa}
\end{align}
Equation \eqref{chiSn} describes the type I process, where a spin-majority quasiparticle is annihilated in the midgap band with the energy $E_\nu<h$ and created in the upper band with the energy $E_\mu >h$.  Equation \eqref{chiSa} describes the type II process, where two quasiparticles with opposite spins are created.

We now substitute the Fourier expansions (\ref{umvm}) into Eqs.~\eqref{chiSn} and \eqref{chiSa} and perform the Fourier transform with respect to $x_1$ and $x_2$.  As discussed in Sec.~\ref{sec:inelastic}, the Fourier momenta $q$ and $K$ correspond to the relative coordinate $x_1-x_2$ and the center-of-mass coordinate $(x_1+x_2)/2$, respectively.  Thus, we obtain the Fourier transforms of the functions $L_{\lambda\mu}^{(\rm I, II)}$ appearing in Eqs.~\eqref{chiS} and \eqref{SzSperp}:
\beg\label{Sk}
\begin{split}
L_{\lambda\mu}^{(\rm I)}(q,K)=\sum\nolimits_{\alpha,\alpha',b,b',\{m_j\}}
& {\cal K}_{\lambda\mu}^{(\rm I)}(\alpha,\alpha',b,b',\{m_j\}) \,
\delta[K-Q(m_1-m_1'+m_2'-m_2)] \\
& \times\delta[p_{\mu,b'}-q-p_{\lambda,b}-\tilde p_F^{(\alpha')}+\tilde p_F^{(\alpha)}
-Q(m_1-m_1'+m_2-m_2')/2],
\\
L_{\mu\lambda}^{(\rm II)}(q,K)=\sum\nolimits_{\alpha,\alpha',b,b',\{m_j\}}
& {\cal K}_{\lambda\mu}^{(\rm II)}(\alpha,\alpha',b,b',\{m_j\}) \,
\delta[K-Q(m_1+m_1'-m_2'-m_2)] \\
& \times\delta[p_{\mu,b'}-q+p_{\lambda,b}-\tilde p_F^{(\alpha')}-\tilde p_F^{(\alpha)}
+Q(m_1+m_1'+m_2+m_2')/2],
\end{split}
\en
where $\{m_j\}=\{m_1,m_1';m_2,m_2'\}$ denotes the set of indices for the Fourier components and
\begin{align}
\label{KI}
{\cal K}_{\lambda\mu}^{(\rm I)}(\alpha,\alpha',b,b',\{m_j\})
& = \left[
\overline{\tilde u}_{\lambda\alpha}^{(b)}(m_1) \tilde{u}_{\mu\alpha'}^{(b')}(m_1')
+\overline{\tilde v}_{\lambda\alpha}^{(b)}(m_1) \tilde{v}_{\mu\alpha'}^{(b')}(m_1')\right]
\left[
\tilde{u}_{\lambda\alpha}^{(b)}(m_2)\overline{\tilde u}_{\mu\alpha'}^{(b')}(m_2')
+\tilde{v}_{\lambda\alpha}^{(b)}(m_2)\overline{\tilde v}_{\mu\alpha'}^{(b')}(m_2')
\right],
\\
{\cal K}_{\lambda\mu}^{(\rm II)}(\alpha,\alpha',b,b',\{m_j\})
& = \frac{1}{2}\left[
\tilde{u}_{\mu\alpha'}^{(b')}(m_1') \tilde{v}_{\lambda\alpha}^{(b)}(m_1)
-\tilde{u}_{\lambda\alpha}^{(b)}(m_1) \tilde{v}_{\mu\alpha'}^{(b')}(m_1')
\right]
\left[
\overline{\tilde u}_{\mu\alpha'}^{(b')}(m_2')
\overline{\tilde v}_{\lambda\alpha}^{(b)}(m_2)
-\overline{\tilde v}_{\mu\alpha'}^{(b')}(m_2')
\overline{\tilde u}_{\lambda\alpha}^{(b)}(m_2)
\right].
\label{KII}
\end{align}
Using Eqs.~\eqref{Swqdef}, \eqref{chiS}, \eqref{KI}, and \eqref{KII}, we obtain Eqs.~\eqref{Sn} and \eqref{Sa}.

\end{widetext}

\section{Technical discussion of the inelastic Bragg scattering}
\label{app:Fourier}


In this Appendix, we present a technical discussion of the elementary processes contributing to the inelastic Bragg scattering introduced in Sec.~\ref{sec:inelastic}.

\begin{figure}
\includegraphics[width=3.1in]{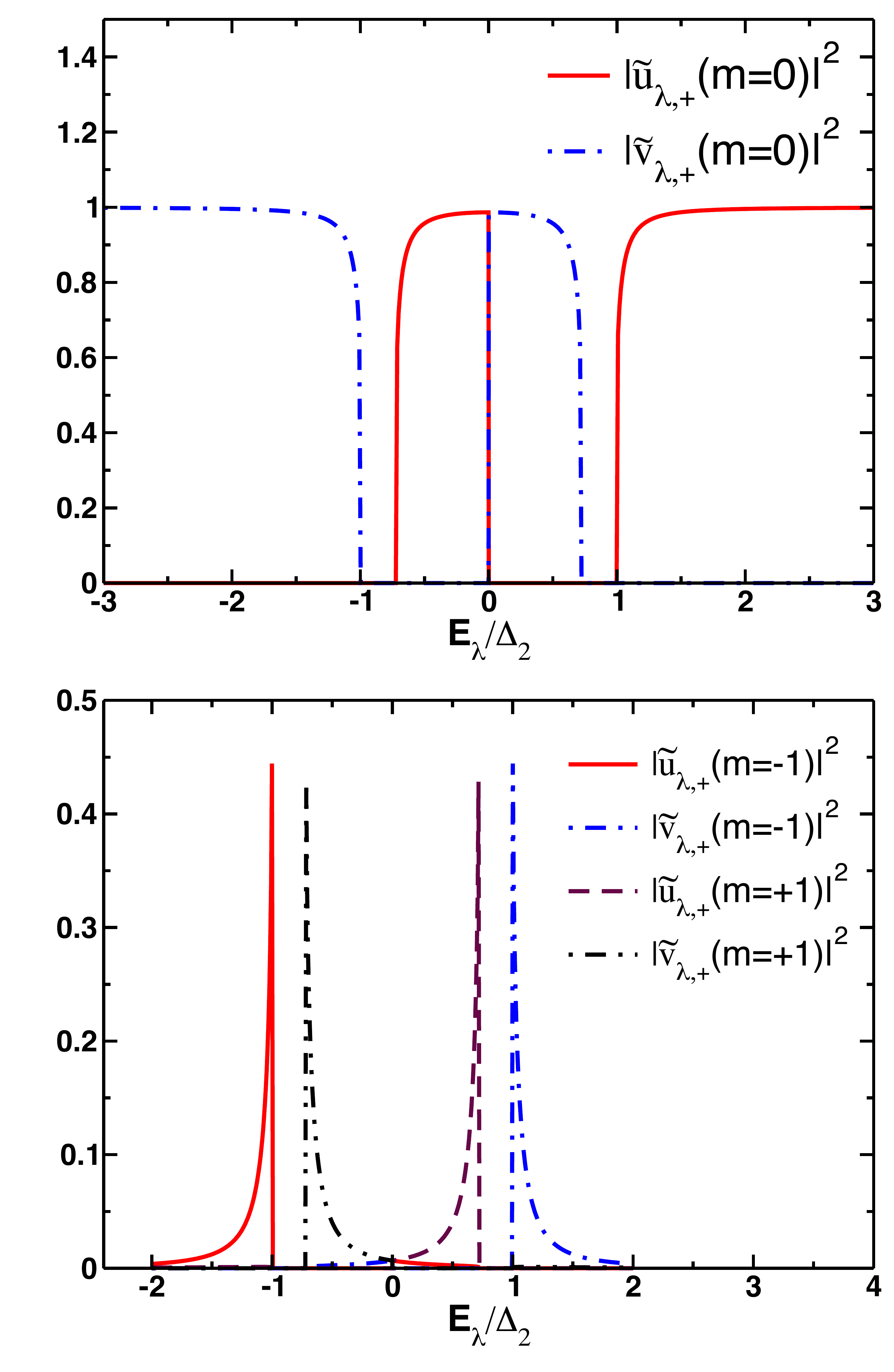}
\caption{(Color online) Plots of the Fourier components of the Bogoliubov amplitudes $|\tilde u_{\lambda b}(m)|$ and $|\tilde v_{\lambda b}(m)|$ from Eq.~(\ref{umvm}) for $m=0,\pm 1$, branch $b=+$, and the right Fermi point ($\alpha=1$).  The amplitudes for the branch $b=-$ can obtained from the ones shown above using $|\tilde u_{\lambda, b}(m)|=|\tilde v_{\lambda,-b}(-m)|$.  The amplitudes for the left Fermi point ($\alpha=2$) can be obtained using Eqs.~\eqref{eq:UnVn}.  Energy dependence of these functions reflects the subtle
structure of the wave function describing the quasi-1D FFLO state.
For the presentation purposes, we have chosen a large effective magnetic field $h=1.95h_c$, along with $k\simeq 0.7$ and $\Delta_2\simeq 1.4\Delta_0$.}
\label{fig:uv}
\end{figure}

\begin{figure}
\includegraphics[width=3.4in]{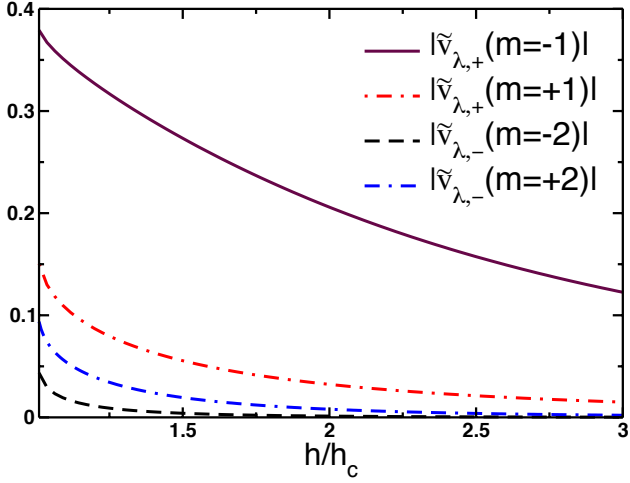}
\caption{(Color online) Plots of the absolute values of the Fourier amplitudes  $|\tilde v_{\lambda,b}(m)|$ vs.\ $h/h_c$ for $E_\lambda=1.2\,{\cal E}_3$.
The higher-order Fourier amplitudes decrease fast with the increase of the effective magnetic field $h$.}
\label{fig:uvmagnetic}
\end{figure}

We begin by discussing the properties of the Fourier transforms of the Bogoliubov amplitudes $\tilde u_{\lambda,b}(m)$ and $\tilde v_{\lambda,b}(m)$ defined in Eq.~\eqref{umvm}.  The absolute values of several Fourier amplitudes $\tilde u_{\lambda,b}(m)$ and $\tilde v_{\lambda,b}(m)$ are plotted in Fig.~\ref{fig:uv} vs.\ the energy $E_\lambda$ for $b=+$.  The particle-hole symmetry relation $|u_{\lambda, b}(m)|=|v_{\lambda,-b}(-m)|$ allows one to eliminate the processes permitted by conservation laws but forbidden by the particle-hole symmetry.  The plots of $|\tilde v_{\lambda, b}(m)|$ vs.\ $h/h_c$ for a fixed value of $E_\lambda=1.2\,{\cal E}_3$ are shown in Fig.~\ref{fig:uvmagnetic}.  We observe that the Fourier components $\tilde u_{\lambda,b}(m)$ and $\tilde v_{\lambda,b}(m)$ are small for $m\not=0,\pm 1$ for the experimentally relevant values of spin polarization, which correspond to the parameter $k\sim 1/2$.  Therefore, we will only discuss matrix elements involving the Fourier amplitudes with $m=0,\pm 1$ .

Now we discuss the origin of lines A and B in Fig.~\ref{fig:formfactor} for the spin structure factor $S(\Omega,q)$ calculated in Sec.~\ref{sec:inelastic}.  First, it is useful to convert the sums over the eigenstates $\lambda$ and $\mu$ in Eqs.~(\ref{Sn}) and (\ref{Sa}) into the energy integrations by introducing the density of states $\rho(\varepsilon)=\sum_\lambda\delta(\varepsilon-E_\lambda)$.  Then, one of the energy integrals can be taken by resolving the delta function representing the energy conservation constraints $E_\mu=\Omega\pm E_\lambda$ in Eqs.~\eqref{Sn} and \eqref{Sa}.  Another energy integral can be taken by resolving the delta function representing momentum conservation:
\begin{align}\label{eq:etados}
S_{\rm I}(\Omega,q) & = \int d\varepsilon \,
\rho_b(\varepsilon)\,\rho_{b'}(\Omega+\varepsilon)\,
{\cal K}^{(\rm I)}(\varepsilon)
\\
&\times\delta[p_{b'}(\Omega+\varepsilon)-q-p_{b}(\varepsilon)+Q(m_1'- m_1)]
\nonumber \\ \nonumber
&=\sum\limits_{\varepsilon_*}\frac{\rho_b(\varepsilon_*)\,
\rho_{b'}(\Omega+\varepsilon_*)\,{\cal K}^{(\rm I)}(\varepsilon_*)}
{|v_{b'}^{-1}(\Omega+\varepsilon_*)-v_b^{-1}(\varepsilon_*)|},
\end{align}
and
\begin{align}\label{eq:etados1}
S_{\rm II}(\Omega,q) & = \int d\varepsilon \,
\rho_b(\varepsilon)\,\rho_{b'}(\Omega-\varepsilon)\,
{\cal K}^{(\rm II)}(\varepsilon)
\\
&\times\delta[p_{b'}(\Omega-\varepsilon)-q+p_{b}(\varepsilon)+Q(m_1'+ m_1)]
\nonumber \\ \nonumber
&=
\sum\limits_{\varepsilon_*}\frac{\rho_b(\varepsilon_*)\,
\rho_{b'}(\Omega-\varepsilon_*)\,{\cal K}^{(\rm II)}(\varepsilon_*)}
{|v_b^{-1}(\varepsilon_*)-v_{b'}^{-1}(\Omega-\varepsilon_*)|},
\end{align}
where summation over $\alpha$, $b$, $b'$, and $\{m_j\}$ is implied, and $v_b^{-1}(\varepsilon)=dp_b(\varepsilon)/d\varepsilon$ is the inverse group velocity of quasiparticles, which is related to the density of states $\rho_b(\varepsilon)=|v_b^{-1}(\varepsilon)|/2\pi$.  The sums in Eqs.~\eqref{eq:etados} and~\eqref{eq:etados1} are taken over the roots $\varepsilon_*$ of the equations representing momentum conservation for the type I and II processes, respectively:
\begin{eqnarray}
\label{eq:pconserv1}
&p_{b'}(\Omega+\varepsilon_*)=q+ p_{b}(\varepsilon_*)-Q(m_1'- m_1), \\
\label{eq:pconserv2}
&p_{b'}(\Omega-\varepsilon_*)=q- p_{b}(\varepsilon_*)-Q(m_1'+ m_1).
\end{eqnarray}
The roots $\varepsilon_*$ exist only for the values of $(q,\Omega)$ located above or at the threshold lines A and B in Fig.~\ref{fig:formfactor}.

The denominators in Eqs.~\eqref{eq:etados} and \eqref{eq:etados1} may vanish for certain combinations of the signs of $b$ and $b'$ when the two branches have equal group velocities $v_b^{-1}=v_{b'}^{-1}$ at the intersection point.  This condition is equivalent to the condition discussed in Sec.~\ref{sec:inelastic} that one energy branch touches another one when displaced by $q$ and $\Omega$.  Vanishing of the denominators in Eqs.~\eqref{eq:etados} or \eqref{eq:etados1} results in divergence of $S(\Omega,q)$ at the corresponding values of $(q,\Omega)$, which constitute lines A and B in Fig.~\ref{fig:formfactor}.  The singularity is smoothed out by a small but finite value of $\delta$ in Eq.~\eqref{Swqdef}.

Now we discuss the contributions of different processes in more detail.  First, we consider the type I processes described by Eqs.~\eqref{eq:etados} and \eqref{eq:pconserv1}, where a quasiparticle with the energy $\varepsilon<{\cal E}_2$ is transferred from the occupied midgap band to the unoccupied upper band with the energy $\varepsilon+\Omega>{\cal E}_3$.  Assuming that $m_1=m_1'=0$ and taking into account the energy spectrum shown in Fig. \ref{fig:exactdispersion}, we observe that the denominator in Eq.~\eqref{eq:etados} can vanish for transitions between the midgap branch with $b=-$ and the upper branch with $b'=+$.  The two energy branches touch for the values of $(q,\Omega)$ belonging to line B in Fig.~\ref{fig:formfactor}, where Eq.~\eqref{eq:etados} has singularity, and Eq.~\eqref{eq:pconserv1} has only one root.  The minimal energy-transfer threshold $\Omega={\cal E}_3-{\cal E}_2$ is achieved at $q=0$.  For the values of $(q,\Omega)$ located above line B in Fig.~\ref{fig:formfactor}, Eq.~\eqref{eq:pconserv1} has two roots, and Eq.~\eqref{eq:etados} gives a non-singular contribution to $S(\Omega,q)$.

There are also the type I excitations with $b'=b$.  As shown in Fig.~\ref{fig:exactdispersion}, such processes require a momentum transfer $Q$ from the lattice, i.e.,\ $m_1'-m_1=\pm 1$.  However, the contribution from these excitations is much smaller than from $b\neq b'$ for the following two reasons: 1) The two group velocities have opposite signs, so the denominator in Eq.~\eqref{eq:etados} does not vanish for $b=b'$; 2) The matrix elements for these transitions with $m_1'-m_1=\pm 1$ quickly decay as a function of energy $\varepsilon$, as shown in Fig.~\ref{fig:uv}.  Thus, the processes with $b=b'$ do not give a significant contribution to $S(\Omega,q)$.

Another contribution to line B in Fig.~\ref{fig:formfactor} comes from the type IIb processes with energies $\varepsilon<{\cal E}_2$ and $\Omega-\varepsilon>{\cal E}_3$.  Given that momentum conservation requires $\alpha\neq \alpha'$ in Eq.~(\ref{KII}), the matrix elements for such processes involve the products $\tilde u_{\mu,b}(m_1')\,\tilde u_{\nu,b'}(m_1)$ and $\tilde v_{\nu,b'}(m_1')\,\tilde v_{\nu,b}(m_1)$.  Similarly to the type I processes, the dominant contribution to the matrix elements comes from $b\neq b'$ and $m_1=m_1'=0$.  However, these processes do not give a significant contribution to $S(\Omega,q)$, because the denominator in Eq.~\eqref{eq:etados1} does not vanish for $b\neq b'$ and $m_1=m_1'=0$.

There is also a contribution to $S(\Omega,q)$ from the type IIb processes with $b=b'$ and $m_1-m_1'=\pm 1$.  Indeed, Fig.~\ref{fig:uv} shows that the matrix elements $\tilde u_{\mu+}(m_1'=1)\,\tilde u_{\nu+}(m_1=0)$ for $E_{\mu}< {\cal E}_2$ and $E_{\lambda}>{\cal E}_3$ are non-zero and strongly peaked at $E_{\mu}$ close to ${\cal E}_2$.  These type IIb processes with $b=b'$ and $\Omega={\cal E}_3+{\cal E}_2$ strongly enhance $S(\Omega,q)$, because the denominator in Eq.~\eqref{eq:etados1} vanishes.  Geometrically, it is a consequence of the peculiar nesting between the midgap and upper branches at $q=Q$ and $\Omega={\cal E}_3+{\cal E}_2$ in Fig.~\ref{fig:exactdispersion}. The strong enhancement of $S({\cal E}_3+{\cal E}_2,Q)$ is indicated by the bright colors in Fig.~\ref{fig:formfactor}.

Next we discuss the type IIa processes, where two quasiparticles are created in the upper band with the energies $\varepsilon>{\cal E}_3$ and $\Omega-\varepsilon>{\cal E}_3$.  These processes are possible at or above line A in Fig.~\ref{fig:formfactor}.  The dominant contribution comes from $b'=b$, because the denominator in Eq.~\eqref{eq:etados1} can vanish in this case.  The energy-transfer threshold is $\Omega=2{\cal E}_3$, as shown by the horizontal dashed lines in Fig.~\ref{fig:formfactor}.  The momentum-transfer threshold can be determined from Eq.~\eqref{eq:pconserv2}: $2p_{b}({\cal E}_3)=q-Q(m_1+m_1')$, where we used $\Omega=2{\cal E}_3$ and $\varepsilon={\cal E}_3$.  Taking into account that $|p_{b}({\cal E}_3)|=Q/2$ and assuming that $m_1=m_1'=0$, we find the momentum threshold at $q=Q$, as shown by the vertical dashed lines in Fig.~\ref{fig:formfactor}.

From momentum conservation, one might expect an excitation line starting at $q=0$ for the type IIa processes with $b'=b$, $m_1=0$, and $m_1'=-1$.  However, the matrix elements $\tilde u_{\mu,b}(m_1')\,\tilde u_{\nu,b'}(m_1)$ and $\tilde v_{\nu,b'}(m_1')\,\tilde v_{\nu,b}(m_1)$ such processes vanish.  Figure~\ref{fig:uv} shows that the diagonal processes with $b=b'$ are allowed only for $m_1=m_1'=0$, whereas the matrix elements for $m_1=0$ and $m_1'=-1$ are zero.

Finally, we consider the type IIa processes with $b\neq b'$.  Given that $p_{+}({\cal E}_3)=-p_{-}({\cal E}_3)$ at the threshold energy, the momentum conservation law reads $q=Q(m_1+m_1')$.  According to Fig.~\ref{fig:uv}, the matrix elements vanish for $m_1=m_1'=0$, but are non-zero for $m_1+m_1'=1$.  Thus, the momentum threshold is $q=Q$ for $b\neq b'$ as well.  However, the processes with $b\neq b'$ give a smaller contribution to $S(\Omega,q)$, because the denominator in Eq.~\eqref{eq:etados1} does not vanish in this case.

\end{appendix}



\begin{thebibliography}{99}
\bibitem{Fulde-note} Fulde and Ferrell \cite{FF} proposed a complex order parameter $\Delta(x)$ proportional to either $\exp(iQx)$ or $\exp(-iQx)$.  This order parameter opens only one of the two energy gaps and, thus, is less favorable than the real one proportional to $\sin(Qx)$.
\end{thebibliography}
\end{document}